\documentclass[a4paper,11pt]{article}
\pdfoutput=1 
\usepackage{jinstpub}
\usepackage{booktabs}
\usepackage{siunitx}
\usepackage{float}
\usepackage{subcaption}

\title{Precise simulation of electromagnetic calorimeter showers using a Wasserstein Generative Adversarial Network}

\author[a]{Martin Erdmann}
\author[a]{Jonas Glombitza}
\author[a,b]{Thorben Quast}

\affiliation[a]{III. Physikalisches Institut A, Rheinisch Westf\"alische Technische Hochschule,\\Aachen, Germany}
\affiliation[b]{EP-LCD, CERN,\\Geneva, Switzerland}

\emailAdd{thorben.quast@cern.ch}

\begin{document}

\begin{abstract}
{Simulations of particle showers in calorimeters are computationally time-consuming, as they have to reproduce both energy depositions and their considerable fluctuations. A new approach to ultra-fast simulations are generative models where all calorimeter energy depositions are generated simultaneously. We use GEANT4 simulations of an electron beam impinging on a multi-layer electromagnetic calorimeter for adversarial training of a generator network and a critic network guided by the Wasserstein distance. The generator is constraint during the training such that the generated showers show the expected dependency on the initial energy and the impact position. It produces realistic calorimeter energy depositions, fluctuations and correlations which we demonstrate in distributions of typical calorimeter observables.
In most aspects, we observe that generated calorimeter showers reach the level of showers as simulated with the GEANT4 program.}
\end{abstract}


\begin{arxivnumber}
{1807.01954}
\end{arxivnumber}

\DeclareRobustCommand{\orderof}{\ensuremath{\mathcal{O}}}
	\maketitle
	\flushbottom
	
    \section{Introduction}\label{sec:Introduction}
    
Calorimeter showers initiated by a primary particle can be understood as a sequence of stochastic interaction processes of the primary and all secondary particles with the material. The four-dimensional spatial and temporal development of the shower and its energy depositions can be described with good accuracy by numerical methods taking into account the relevant cross sections, e.g., using the GEANT4 program \cite{Agostinelli:2002hh,Allison:2006ve,Allison:2016}. Sequential simulations of particle showers, however, require significant computing resources.

The most recent approach for simulating particle showers in calorimeters are so-called generative adversarial networks (GAN) \cite{2014arXiv1406.2661G,2016arXiv161207828S,Hooberman:2017nips,Paganini:2017hrr,Paganini:2017dwg}. According to this concept, the temporal sequence of the shower development is first marginalized by training a generator and the three-dimensional spatial distribution of the energy depositions is then generated directly. At first glance, this ansatz appears similar to common detector-specific parameterizations of detailed simulations which are usually developed by experts in the field. With the GAN concept, however, a high-dimensional probability distribution for spatial energy depositions is obtained automatically  either directly from measured data, or alternatively from the above-mentioned detailed simulations.

In order to produce libraries of network-generated particle showers with the relevant properties of measured showers, the probability distribution of realistic energy depositions needs to be encoded in the numerous trainable parameters of the network. The key challenge is to build a converging framework which is capable of approximating the entire high-dimensional probability distribution.

In contrast to the above-mentioned sequential shower simulations, which calculate many stochastic processes of individual interactions, the entire spatial energy depositions of the shower are determined in a single evaluation of the generator network. In order to obtain the realization of a single shower, the stochastic process is incorporated through a set of random numbers on input to the probability distribution coded in the network. These random numbers ensure that none of the generated particle showers look alike. The speed for calculating a shower realization is several orders of magnitude faster than detailed shower simulations, since in the network only a fixed sequence of linear algebra operations is carried out together with the evaluation of the activation functions.

Current research is assessing methods of training the generator to learn the high-dimensional probability distribution for calorimeter energy depositions. The GAN concept consists of two networks working in opposition to one another. The generator network is meant to learn the probability distribution which is encoded in realistic data sets. The second network is used to evaluate the differences between the generated data sets and the realistic data sets. The feedback of the second network to the generator network is used to improve the probability distribution encoded in the generator. Conversely, the second network is trained to distinguish between ever smaller remaining differences between the generated data sets and the realistic data sets. In this dual training process, the probability distribution of energy depositions is sampled from realistic data and transferred to the generator network.

In the original work on GANs for simulating particle showers in calorimeters, the evaluating network was a binary classifier that returned as feedback a probability for each shower to be real. So far it turned out that here the generator training was only partially successful.

Instead of using binary classification in the evaluating network, recently a high-dimensional distance measure between the probability distributions of example data and generated showers was successfully used for simulating an atmospheric calorimeter with a single readout layer \cite{Erdmann:2018kuh}. In this simulation, the adversarial training was performed using the Wasserstein distance as it was applied in computer science research in the past \cite{2017arXiv170107875A,2017arXiv170400028G}. This variant of the GAN concept is referred to as WGAN (Wasserstein GAN).

In this paper we apply the WGAN concept to the concrete setup of a test beam for a realistic electromagnetic sampling calorimeter, which consists of several pixelated readout layers interspersed with absorbers. Primary beam particles are electrons with different energies $E$ impinging perpendicularly at different positions $(P_x,P_y)$ on the front surface of the calorimeter. To generate dedicated particle showers for the initial conditions $(P_x,P_y,E)$ of an electron, the architecture of the WGAN is supported by two additional networks constraining these beam conditions. The concept of using conditioning networks together with generative adversarial networks has been explored before \cite{2014arXiv1406.2661G,Hooberman:2017nips,Paganini:2017hrr,Erdmann:2018kuh}.

The requirements for the generation of electromagnetic calorimeter showers are considerable. In each layer of the calorimeter, the transverse energy depositions must correspond to those of a particle shower. For the description of the longitudinal shower development, layer-wise correlations of reconstructed observables are also decisive. Therefore, a considerable proportion of our study is devoted to the quality assessment of the generated showers. This concerns aspects that are directly trained for, such as the initial beam conditions, as well as physical aspects which have to be learned during the adversarial training.

Our paper is structured as follows. We start by presenting the test beam experiment for the calorimeter and describe the data simulations used as sample data sets for training the WGAN. After that we explain the network structure of the WGAN together with the usage of the Wasserstein distance and the constrainer networks to respect the beam conditions. We then examine in detail the quality of the generated showers by comparing in particular the generated calorimeter showers with showers simulated using the GEANT4 program. Finally, we present our conclusions.

\section{Experimental setup}\label{sec:HGCal}
\subsection{Calorimeter configuration}\label{subsec:HGCal}
\begin{figure}{b}
	\centering
	\includegraphics[width=0.35\textwidth]{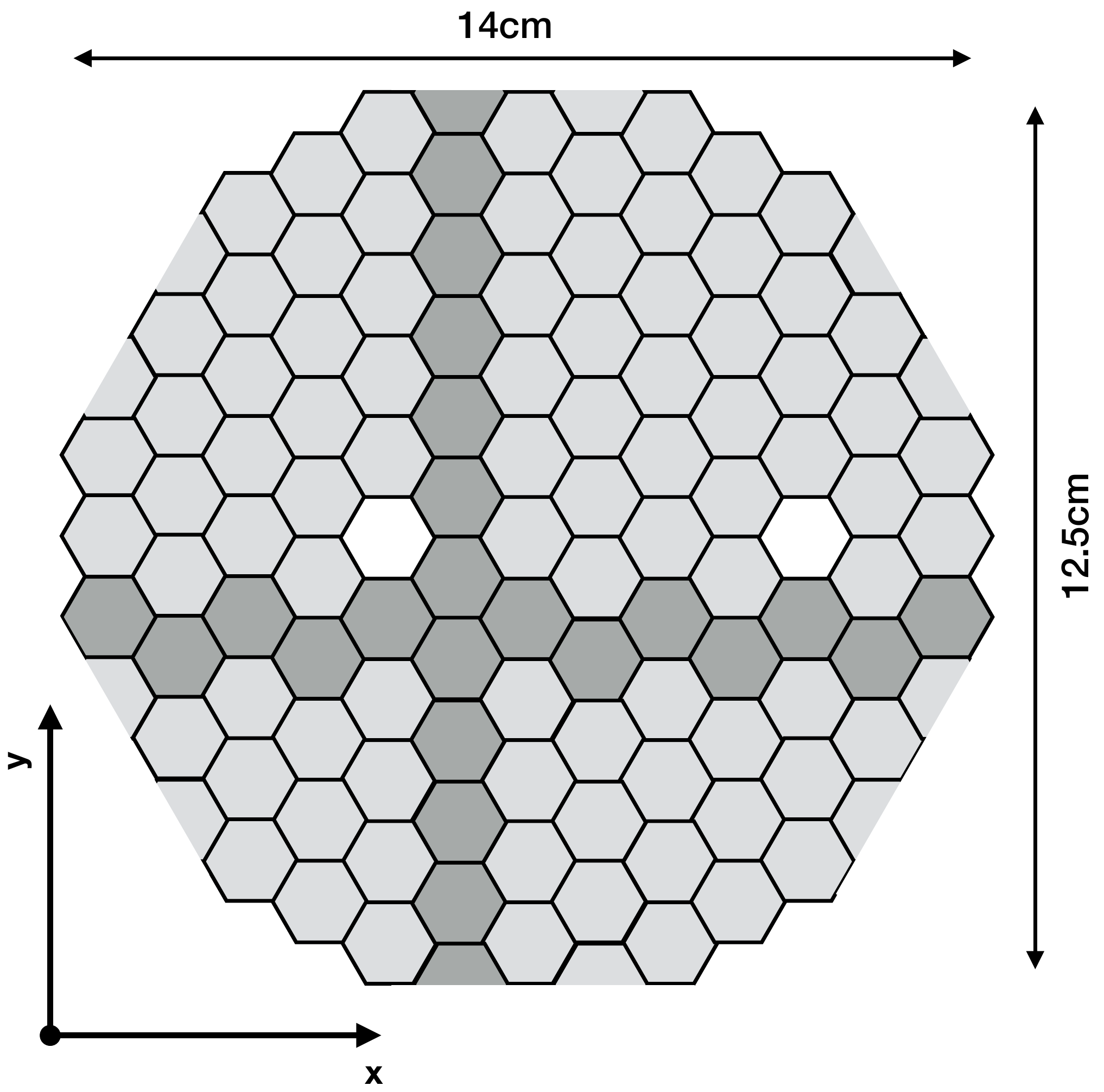}
	\caption{Full wafer pixelation. Shaded pixels are of constant $x$ and $y$ coordinates, respectively.}
	\label{fig:CoordinateMapping}	
\end{figure}
	\begin{figure*}{b}
	\centering 
	\includegraphics[width=0.7\textwidth]{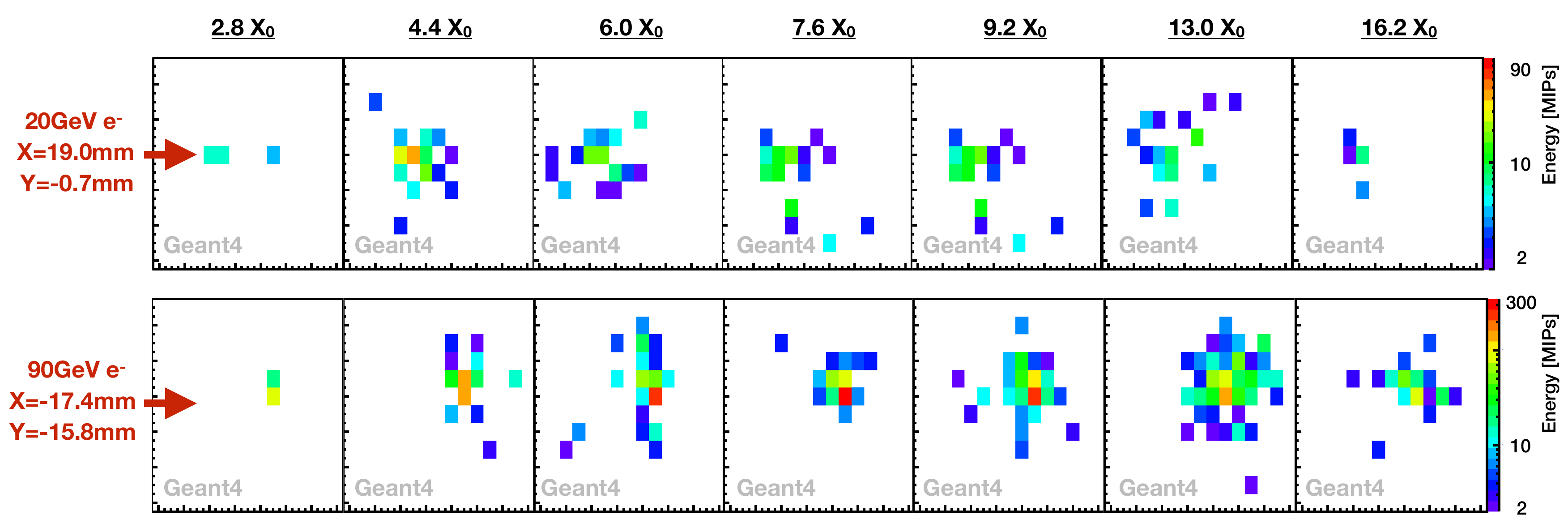}
	\caption{Energy depositions of an electromagnetic shower induced by a $20~$GeV electron (top) and $90~$GeV electron (bottom) with different impact positions (X,Y) simulated using GEANT4. The 3D shower images consist of 12$\times$15$\times$7 pixels.}
	\label{fig:GEANT4Displays}
\end{figure*}
To discuss generative shower models in the context of a realistic detector, we choose from many possible calorimeter setups a configuration of the electromagnetic compartment of a CMS High Granularity Calorimeter (HGCAL) prototype \cite{CMS:2008CMSExperiment,Contardo:2018TechnicalReportHGCal,Martelli:2018HGCalOverview}.      
Various other HGCAL prototypes with different sampling configurations have already been tested with beams at CERN and Fermilab since 2016 \cite{Jain:2017BeamtestSummary2016,Quast:2018BeamtestSummary2017}.		
This specific configuration incorporates seven sensitive layers covering 2.8 - 16.2 radiation lengths ($X_{0}$) of electromagnetic showers and was tested with highly energetic secondary electrons at CERN's Super Proton Synchrotron test beam facility in September 2017.   
Each sensor is made of one 6-inch hexagonal silicon n-type wafer. Its active thickness at full depletion amounts to 300$~\mu$m. Most of the 135 individual pixels on each wafer are 1$~$cm$^2$-sized hexagons. Pixels at the edges have various shapes. Only full and half pixels at the edges are considered in this study. Furthermore, centrally placed calibration pixels are treated as dead pixels here and hence are excluded from the shower measurement. Instead of using directly hexagonal geometries, a coordinate system is constructed such that the pixel positions are indicated in a 12$\times$15 Cartesian-like frame following \cite{Erdmann:2018kuh}. Shaded hexagons in Figure \ref{fig:CoordinateMapping} illustrate the lines of constant $x$ and $y$ coordinates, respectively.
	After their assembly to full modules, the wafers are glued to 1.2$~$mm thick copper-tungsten baseplates and subsequently inserted into a hanging file system where they are interspersed by 6$~$mm thick copper- and 4.9$~$mm thick iron-coated lead absorbers.
	The test beam line is 15$~$m long and adds another 0.27$~X_{0}$ of upstream material. It comprises six gas-filled delay wire chambers (DWC) \cite{Spanggaard:1998H2DWCs} and six scintillation counters.
		
	\subsection{Reference dataset}\label{subsec:dataset}

	The training and the subsequent evaluation of the WGAN performance require a well-defined reference dataset of electromagnetic showers. In general, this set is sampled from an underlying highly dimensional probability density which the WGAN is ultimately supposed to learn. Showers taken from this dataset are referred to as "real" in the following. \newline
	In this paper, we construct sequences of real showers from simulation of electromagnetic cascades with GEANT4 version 10.2 \cite{Agostinelli:2002hh,Allison:2006ve} using a specific tune of the FTFP\_BERT physics list \cite{Banerjee:2017CMSPhysicsLists}. The geometry of the calorimeter and of the test beam line is implemented within a release of the official CMS offline computing software which is publicly available on GitHub \cite{github}. 
	Similar to real test beam data, energy depositions are converted into units of signal produced by minimum ionizing particles (MIPs) traversing a pixel. \newline Furthermore, in order not to rely on idealized assumptions in the simulation, various constraints of typical beam tests of such a calorimeter are taken into account: 
	\begin{itemize}
		\item Electrons inducing showers traverse the upstream material in the beam line and impinge perpendicularly onto the calorimeter.
		\item Impact positions are extrapolated from straight line tracks computed from four position measurements as they would be measured by four DWCs in the beam line with 200$~\mu$m resolution each.						
		\item The beam profile is modeled with a rectangular geometry and covers an active area of 6$\times$5$~$cm$^2$. Its energies are smeared with a 1\% uncertainty.
		\item Calorimeter pixels with energy depositions below 2 MIPs are removed from each shower to reject noise contributions.
	\end{itemize}
	The dataset for training the WGAN consists of $5\times 100,000$ electromagnetic showers induced by $20~$GeV, $32~$GeV, $50~$GeV, $80~$GeV and $90~$GeV electrons. Due to the smearing of the energy and impact position measurements, the assigned labels deviate from the nominal values at the percentage level. By way of example, Figure \ref{fig:GEANT4Displays} illustrates two real showers induced by one $20~$GeV and one $90~$GeV electron, respectively.

	For further evaluation, another dataset is constructed from simulated $70~$GeV electrons. This sample serves exclusively to investigate the WGAN's interpolation capacities to energy labels for which it has not been trained. The aforementioned beam energy smearing of 1\% is not applied for this particular dataset.

	\section{Fast simulation approach}\label{sec:Fastsim}    	
	\subsection{Generative adversarial networks (GANs)}\label{subsec:GAN}
	Generative adversarial networks (GANs) are a widely used concept of generative models that was introduced by Goodfellow et al. in 2014 \cite{2014arXiv1406.2661G}.
	The framework consists of two adversarial networks, namely a generator network $G$ and a discriminator network $D$. The overall goal of this adversarial framework is to train a generator to be able to generate samples $\tilde{x} = G(z)$ out of noise $z$, which are very similar to real samples $x$.
	During the training process the generator improves its performance using the feedback provided by the discriminator, which measures the similarity between generated and real samples.
	
	Even though traditional GANs show impressive results, the training process is unstable and hard to monitor. Furthermore, GANs often suffer from mode collapsing when the generator is only able to generate data in a subspace of the real distribution.
	The recently published Wasserstein GAN \cite{2017arXiv170107875A} and its improvement \cite{2017arXiv170400028G} allows for a stabilized training procedure by delivering adequate gradients to the generator, providing a meaningful loss metric not being susceptible to mode collapsing.
	In the following section, we first introduce the Wasserstein GAN and the method of label conditioning, and then present our network architecture and our training strategy. Finally, we describe the training of our adversarial framework to generate calorimeter showers.
	
	\subsection{Wasserstein GANs}\label{subsec:WGAN}
	In Wasserstein GANs, the Wasserstein-1 metric is used as a similarity measure between the generated samples $\tilde{x} = G(z)$ and the real samples $x$. This distance is also known as Earth mover's distance, because in a figurative sense it defines the cost for moving a distribution onto a target distribution using optimal transport. 
	In the adversarial framework the Wasserstein loss is constructed using the Kantorovich-Rubinstein duality:
	\begin{equation}
	L = \sup_{f \in \mathrm{Lip}_1} \left( \mathbb{E}[ f(x) ] - \mathbb{E}[ f(\tilde{x}) ] \right).
	\label{eq:DW}
	\end{equation}
	Here, "$\sup_{f \in \mathrm{Lip}_1}$" states that the supremum is over all the 1-Lipschitz functions $f$ after application on the real samples $x$ and generated samples $\tilde{x}$.
	During the adversarial training, the 1-Lipschitz functions $f$ which fulfill (\ref{eq:DW}) are approximated by the discriminator network $D$. It is called \textit{critic} because it is trained to allow for an estimate of the Wasserstein distance instead of being able to discriminate between real and generated samples.
	To allow for the approximation of the 1-Lipschitz functions using a neural network, the Lipschitz constraint is enforced by the gradient penalty \cite{2017arXiv170400028G} which extends the objective function to:
	\begin{equation}
	\label{eq:wasserstein_loss}
	L = \mathbb{E}[ D(x) ] - \mathbb{E}[ D(G(z)) ] - \lambda \; \mathbb{E}[(\vert\vert \nabla_{\hat{u}} f_w(\hat{u}) \vert\vert_2 - 1 ) ^2 ]\;.
	\end{equation}
	Here, $\lambda$ is a hyperparameter for scaling the gradient penalty.
	The mixture term 
	\begin{equation}
	\hat{u} = \varepsilon x + (1-\varepsilon) \tilde{x}
	\end{equation}
	states that the Lipschitz constraint is enforced by sampling on straight lines between pairs of generated samples $\tilde{x}$ and real samples $x$. The random sampling is performed by sampling $\varepsilon$ from a uniform distribution $\mathcal{U}(0,1)$. 
	To ensure accurate gradients for the generator, the critic is usually trained for several iterations before one generator update is applied.
	Thus, in Wasserstein GANs the generator attempts to \textit{minimize} the Wasserstein distance (\ref{eq:wasserstein_loss}) between the generated and the real samples, while the Wasserstein distance is approximated using the critic network by \textit{maximizing} (\ref{eq:wasserstein_loss}).
	This differs to the traditional GAN setup where, under the assumption of an optimal discriminator, the generator attempts to minimize the Jensen-Shannon divergence.

	\subsection{Label conditioning}\label{subsec:conditioning}
	
	For calorimeter simulations, generated samples must reflect certain label characteristics according to physics laws. The labels define the initial state of the simulation such as the incident particle's kinematics and the degree of possible background activity (pileup) in the calorimeter.
	However, this label dependency is not ensured for samples generated by a generator which is trained using the WGAN approach.
	To be able to generate samples which can be associated with explicit labels, the concept of label conditioning introduced by Auxiliary Classifier GANs (AC-GANs) \cite{2016arXiv161009585O} is a widely used concept for generative approaches in physics simulations \cite{Hooberman:2017nips,Paganini:2017hrr,Erdmann:2018kuh}.
	To advance the WGAN concept to label conditioning we adapt the concept of \cite{Erdmann:2018kuh}. 
	In this specific configuration, the initial state is determined by the electron energy and its impact position.	Accordingly, the generator dependency is modified to $G=G(z,E,P)$, and in our setup, besides noise $z$, the generator is given the physics labels of the electron energy $E$ and the impact position coordinates $P=(P_x,P_y)$ as input. Furthermore, we also provide the critic with the label information.
	
	To constrain the generator and to evaluate how well the label characteristics are reflected in the generated samples, two constrainer networks $a_i$ are used. These constrainer networks are trained under supervision to reconstruct the impact position and the electron energy respectively using the real (labeled) samples.
	The mean squared error
	\begin{equation}
	L_\mathrm{{real,i}} = [y_i  -  a_i(x)]^{2}
	\end{equation}
	is used as an objective function for the constrainer networks.
	Here, $y_i$ is one label associated with the real sample $x$ and $a_i(x)$ denotes the respective reconstruction by the constrainer network. The constrainer networks are trained under supervision during the critic training and are fixed during the generator training.
	To enforce label conditioning of the generator, the generator loss is extended by
	\begin{equation}
	\label{eq:aux_loss}
	L_{\mathrm{aux}}=\sum_i^n \kappa_i |L_\mathrm{{real,i}}-L_\mathrm{{fake,i}}|,
	\end{equation}
	where $\kappa_i$ is a hyperparameter to scale the respective auxiliary loss. The loss \begin{equation}
	L_\mathrm{{fake,i}} = [\alpha_i - a_i(\tilde{x})]^{2} = [\alpha_i - a_i(G(z, E, P))]^{2},
	\end{equation}
	states how well the input labels $\alpha_1 = E,\; \alpha_2 = P$ for the generation can be reconstructed from the generated shower by the constrainer networks. In summary, the generator is trained to minimize the Wasserstein distance (\ref{eq:wasserstein_loss}) and the auxiliary loss (\ref{eq:aux_loss}) provided by the constrainer networks. The absolute difference between both loss terms in (\ref{eq:aux_loss}) ensures that the label reconstruction of the generated and real samples remains on the same scale.

	\subsection{Strategy and network training}\label{subsec:training}
		\begin{figure}
		\captionsetup[subfigure]{aboveskip=-1pt,belowskip=-1pt}
		\begin{centering}
			\begin{subfigure}[b]{0.5\textwidth}
				\includegraphics[trim={1.8cm 0cm 0.5cm 0cm},clip,,width=\textwidth]{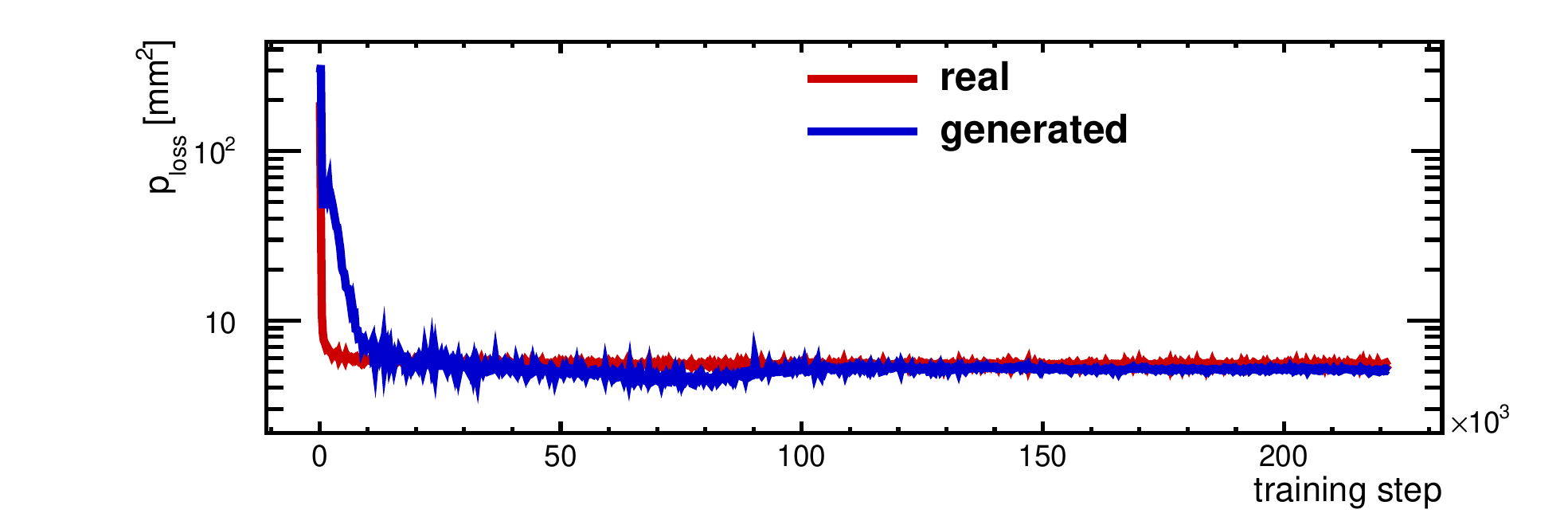}
				\subcaption{}
				\label{fig:cost_p}
			\end{subfigure}
			\hfill
			\begin{subfigure}[b]{0.5\textwidth}
				\includegraphics[trim={1.8cm 0cm 0.5cm 0cm},clip,,width=\textwidth]{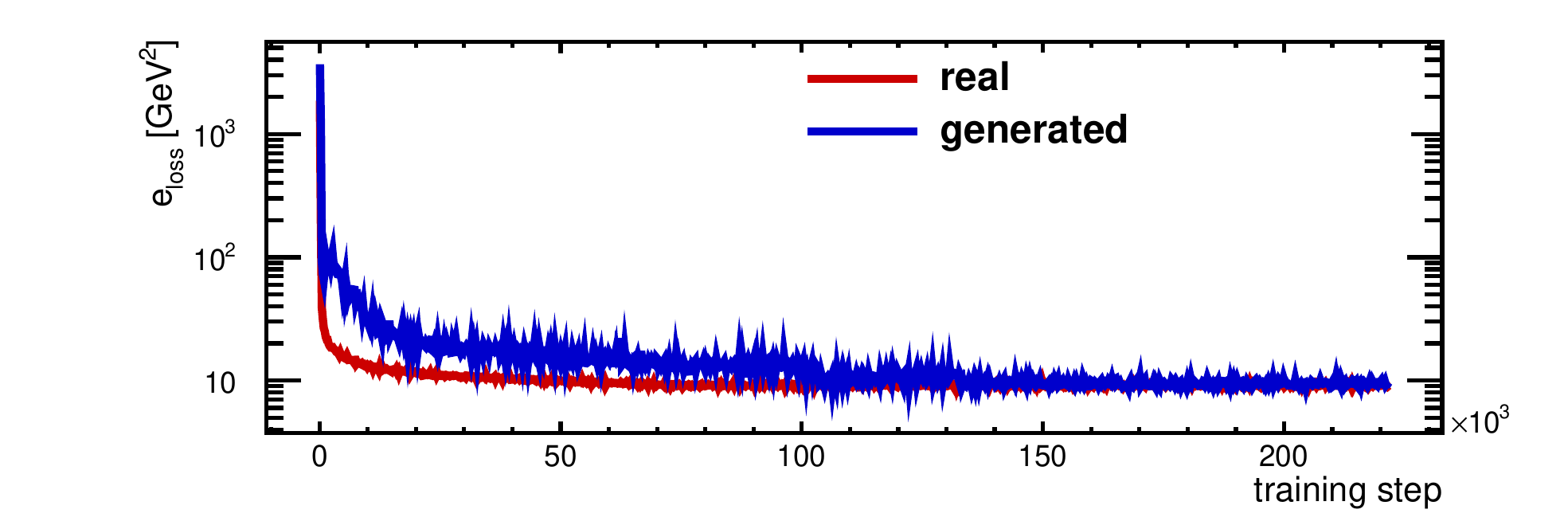}
				\subcaption{}
				\label{fig:cost_e}
			\end{subfigure}
			\hfill
			\begin{subfigure}[b]{0.5\textwidth}
				\includegraphics[trim={1.8cm 0 0.5cm 0cm},clip,,width=\textwidth]{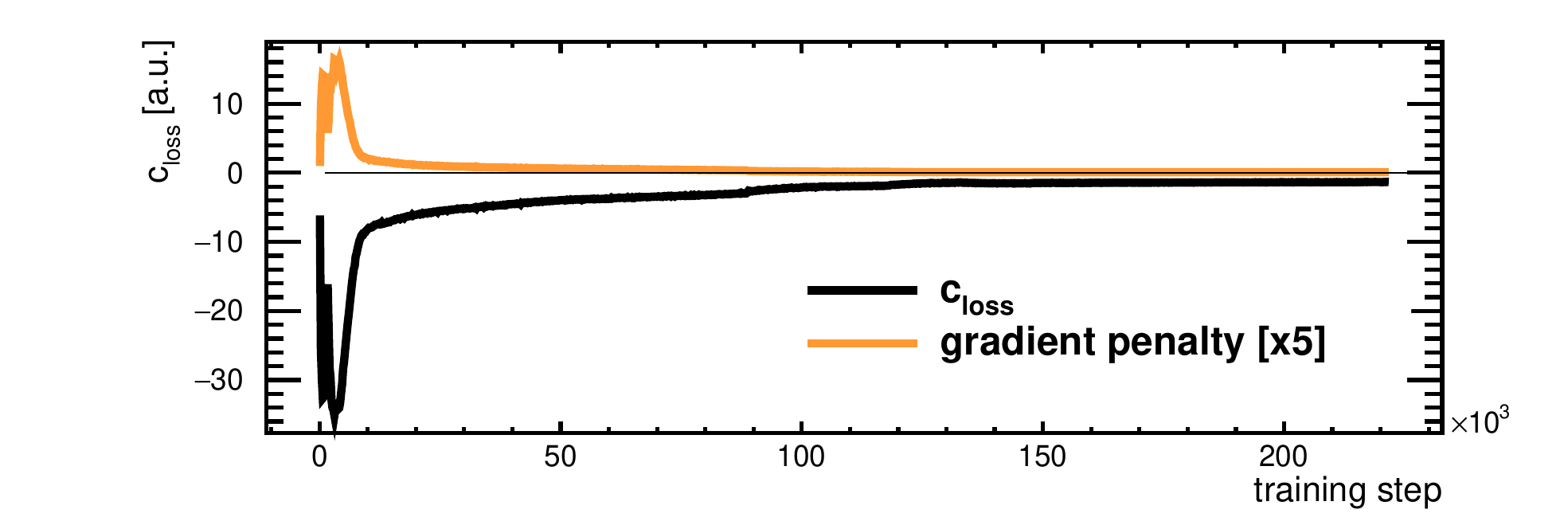}
				\subcaption{}
				\label{fig:cot_c}
			\end{subfigure}
			\caption{Loss curves of the constrainer networks during the supervised training for (a) the position regression and (b) the energy reconstruction. (c) Total critic loss $c_{loss}$ during the training (black) and rescaled gradient penalty (yellow).}
			\label{fig:costs}
		\end{centering}
	\end{figure}	
	Our framework for the generation of electromagnetic calorimeter showers consists of four networks: one generator, one critic and two constrainer networks. One of the constrainer networks is used for conditioning the energy $E$, while the second is used for conditioning the impact position $P$. The networks, their training and their evaluation are implemented using the Tensorflow \cite{tensorflow} framework (v1.5). Exact details of all architectures can be found in the appendix \ref{sec:Appendix} in table \ref{table:critic}, \ref{table:generator} and \ref{table:constrainer}. 
	
	The generator consists of two parts: The first part is separated into 7 towers, each of which has the same structure, and a joint part which merges the towers. Each of the 7 towers is given $10$ latent variables $z$ and $3$ labels $\alpha$ describing the energy and the impact position of the calorimeter shower as input. After two fully connected layers and a reshape, a block of three 2D transposed convolutions and a single 2D convolution follows. Next, the 7 towers are concatenated to a joint part with three 2D convolutional layers. Finally a locally connected convolutional layer completes the generator architecture. Between the convolutional and transposed convolutional layers we use batch normalization and leaky ReLUs as activations. After the last layer we do not apply batch normalization and use ReLU as activation to allow for the generation of sparse calorimeter images. To enlarge the prior for the generation process, a masking layer masks the dead pixels and regions outside the calorimeter by setting the respective values to zero.

		\begin{figure*}
		\centering 
		\includegraphics[width=0.65\textwidth]{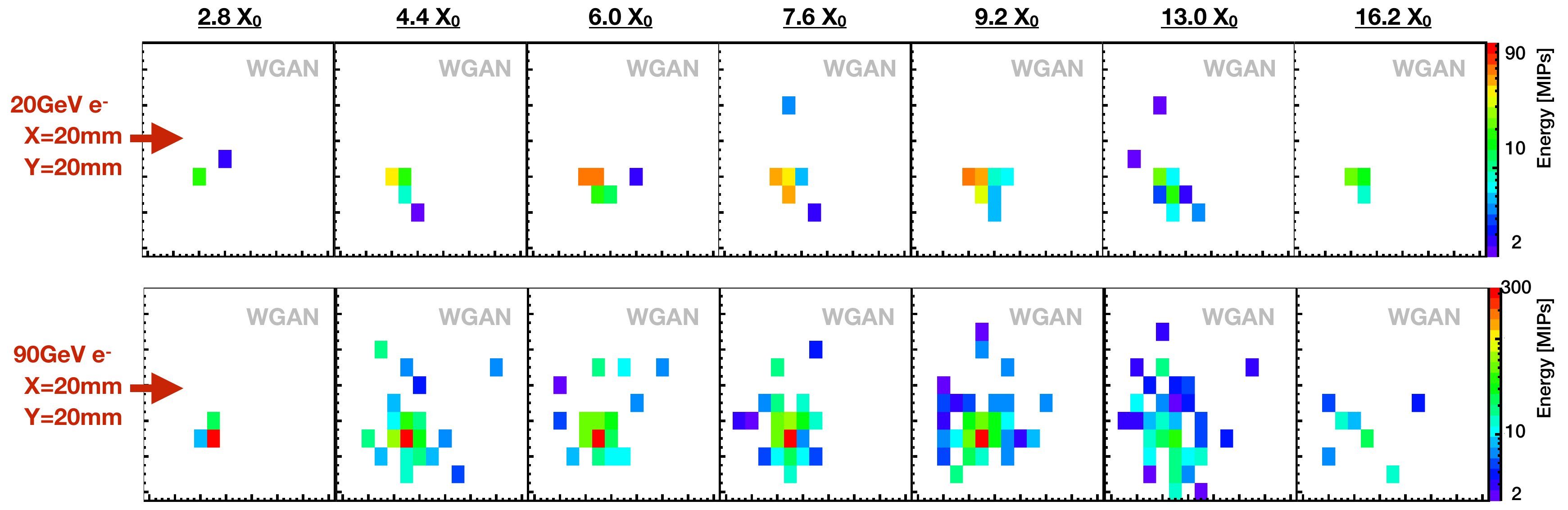}
		\caption{Energy depositions generated with the WGAN for a fixed impact position of an electromagnetic shower for a 20$~$GeV electron (top) and for a $90~$GeV electron (bottom).}
		\label{fig:eventDisplayWGAN}
	\end{figure*}
	\begin{figure*}
		\centering 
		\includegraphics[width=0.65\textwidth]{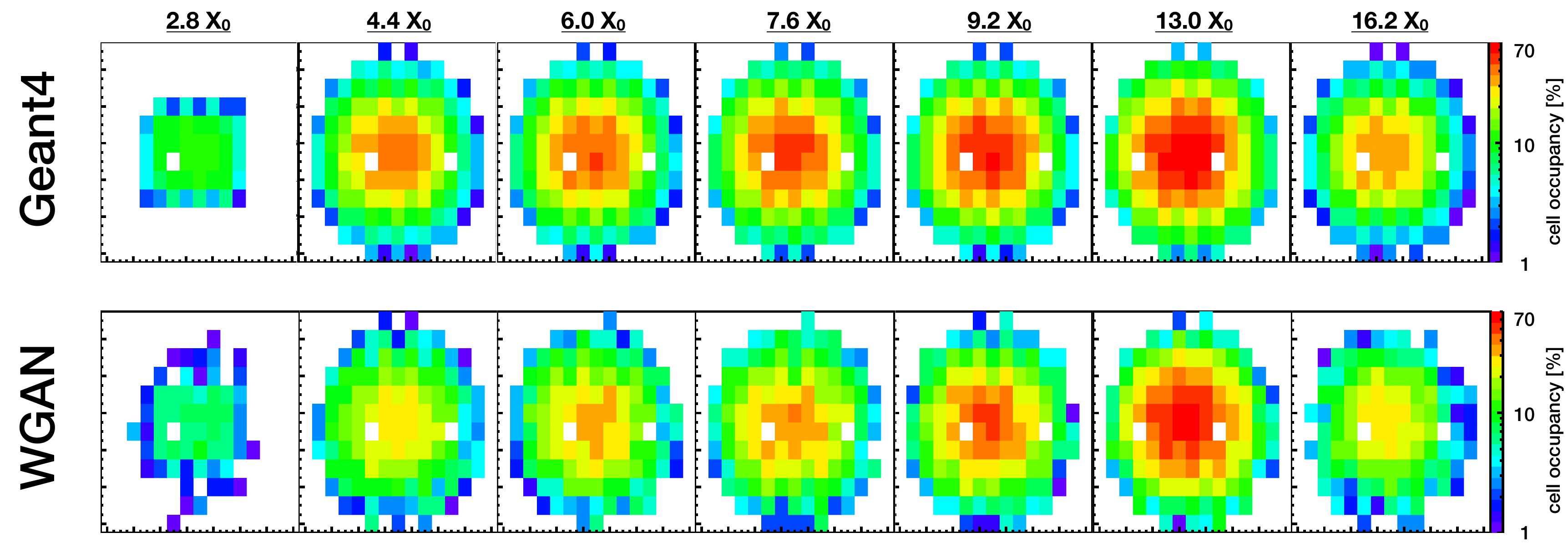}
		\caption{Cell occupancy for $90~$GeV electrons simulated using GEANT4 (top) and generated by the WGAN (bottom). Dead pixels and areas outside the sensor acceptance are masked in the generator.}
		\label{fig:AverageOccupancy}
	\end{figure*}

	The critic network is given as additional input the $3$ labels, which are processed by two fully connected layers and a reshape to obtain a two-dimensional shape.
	The following architecture of the critic is straightforward and consists of five 2D convolutional layers followed by a fully connected layer and the output layer. As activation we use leaky ReLU to avert sparse gradients. Between the layers we use layer normalization instead of batch normalization as we use the gradient penalty loss.
		
	For both constrainer networks we used a very similar architecture of 3D convolutions where we varied only the classification layer. For better convergence and regularizing effects we use batch normalization between the layers. Furthermore, we use leaky ReLU as nonlinearity to ensure sufficient gradients.

	During training the losses of the constrainer networks are scaled with $\kappa_{\mathrm{E}}=\kappa_{\mathrm{P}}=0.01$. The gradient penalty scale is set to $\lambda=5$. We update the constrainer networks and the critic for $n_{cr}=9$ iterations before updating the generator once. We use a batch size of 256 and train the framework for $150$ epochs on a single NVIDIA GeForce GTX 1080 which takes about $30$ hours. We use $10$ latent variables each following a uniform distribution $\mathcal{U}(-1,1)$. Furthermore, we use the Adam optimizers with  $\beta_1=0.0,\;\beta_2=0.9$ \cite{2017arXiv170400028G} and different learning rates for the networks. The constrainer networks use a small learning rate of $lr = 5\cdot 10^{-5}$. Their training is stopped after 50 epochs. For the generator we use a learning rate of $lr = 10^{-3}$ and drop the learning rate after 70, 90 and 100 epochs to $lr = 5\cdot 10^{-4}$, $lr = 2\cdot 10^{-4}$ or rather $lr = 10^{-4}$. For the critic we use an initial learning rate of $lr = 5\cdot 10^{-4}$ and change the learning rate to $lr = 2\cdot10^{-4}$, $lr = 10^{-4}$ and $lr = 5 \cdot 10^{-5}$ after 60, 80 and 100 epochs, respectively.

	\section{Performance benchmarks}\label{sec:benchmarks}
	Various benchmarks related to the quality of our generated electromagnetic showers are discussed in the following. It will be demonstrated that WGAN produces high-quality showers which resemble the real dataset in many aspects while lowering the computing time to simulate a full electromagnetic shower by three orders of magnitude.

	\begin{figure*}
		\captionsetup[subfigure]{aboveskip=-1pt,belowskip=-1pt}
		\begin{centering}
			\begin{subfigure}[b]{0.32\textwidth}
				\includegraphics[trim={0cm 0 1.6cm 0cm},clip,,width=\textwidth]{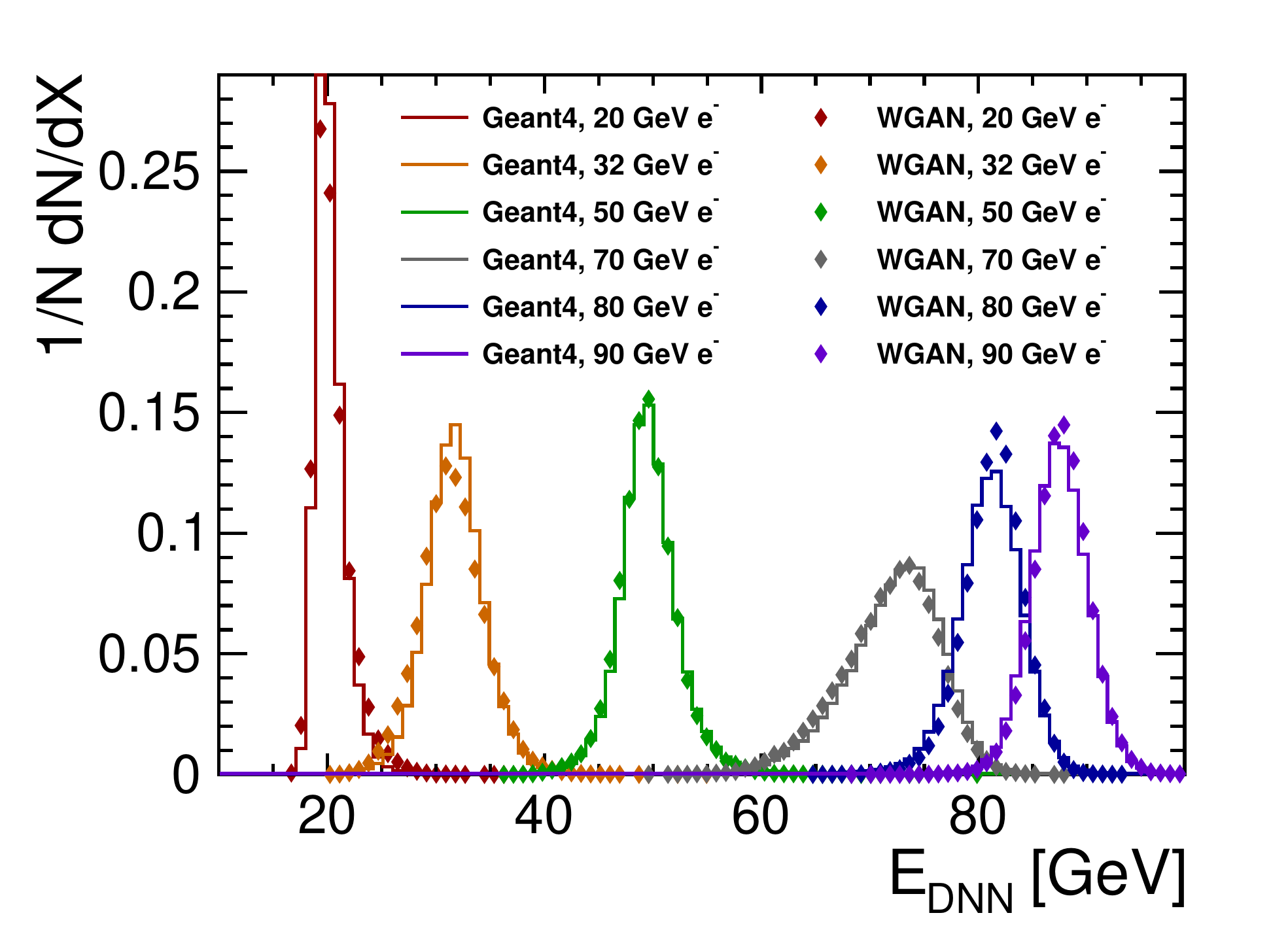}
				\subcaption{}
				\label{fig:dep_energy}
			\end{subfigure}
			\hfill
			\begin{subfigure}[b]{0.32\textwidth}
				\includegraphics[trim={0cm 0 1.6cm 0cm},clip,,width=\textwidth]{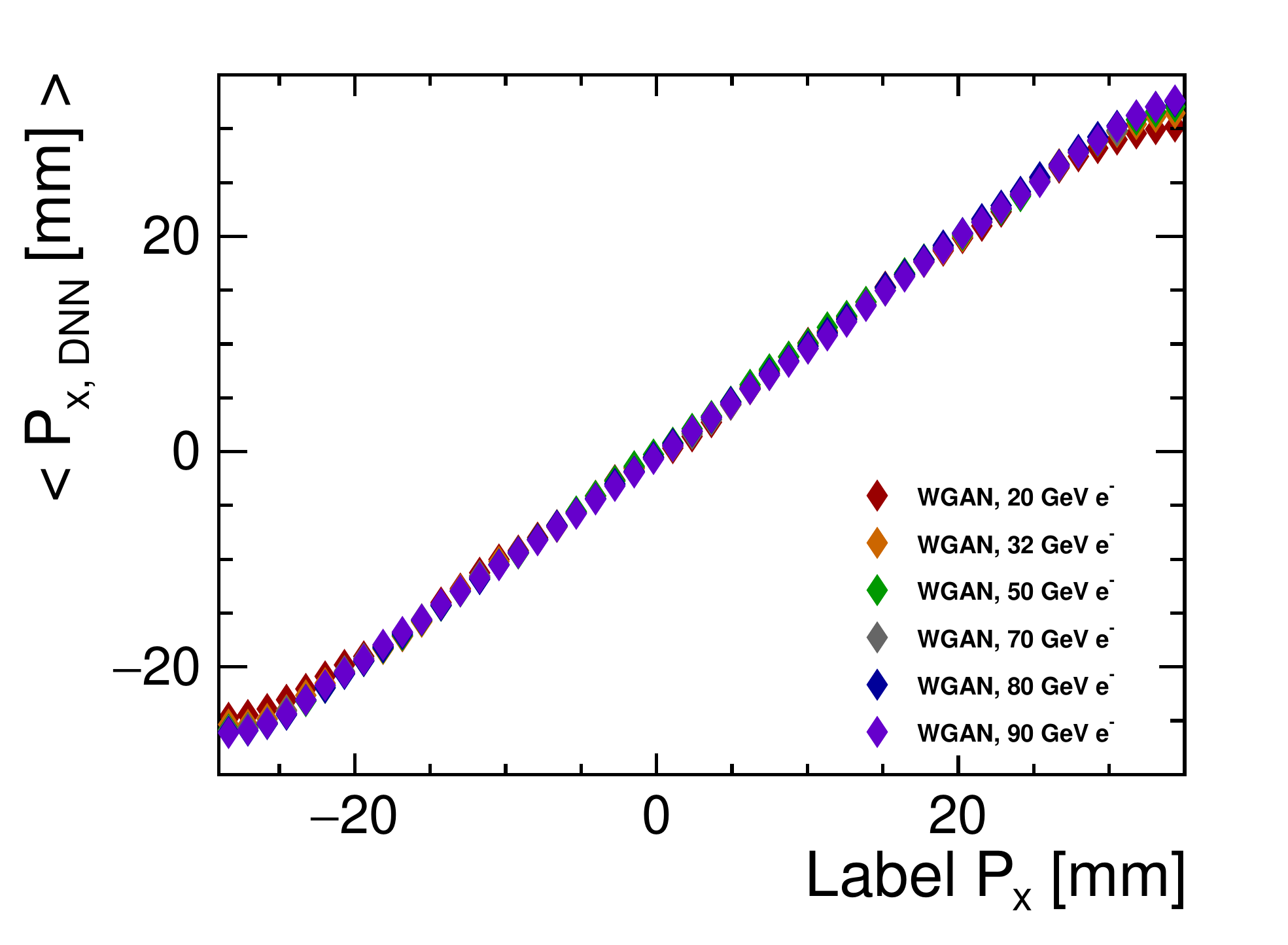}
				\subcaption{}
				\label{fig:dep_posX}
			\end{subfigure}
			\hfill
			\begin{subfigure}[b]{0.32\textwidth}
				\includegraphics[trim={0cm 0 1.6cm 0cm},clip,,width=\textwidth]{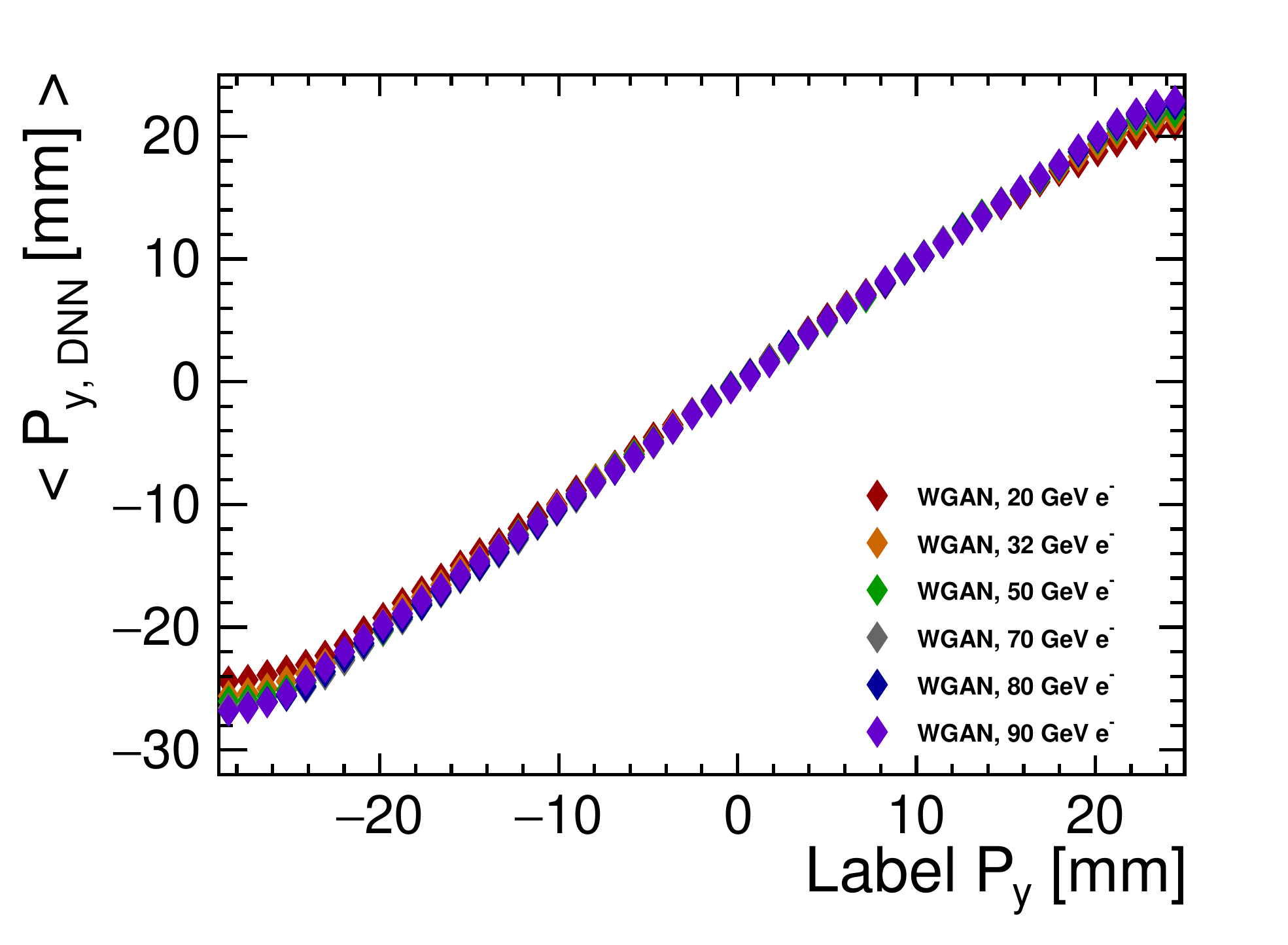}
				\subcaption{}
				\label{fig:dep_posY}
			\end{subfigure}
			\caption{Distribution of energies reconstructed by the constrainer network computed on generated and GEANT4 simulated showers for different energy labels (a). Impact position $P_x$ (b), and impact position $P_y$ (c) of generated showers reconstructed by the constrainer network as a function of their true labels. Statistical errors on these means are negligible. Note that $70~$GeV showers were not part of the training set.}
		\end{centering}
	\end{figure*}   
    
	The investigation is structured as follows. In section \ref{subsec:displays} we perform a visual inspection of WGAN generated showers. First, we illustrate two examples on which qualitative observations are highlighted. Second, the analysis of pixel occupations reveals that the trained WGAN considers the radially decreasing occupancy profile.
	In section \ref{subsec:labels} we then show that the generated samples reflect characteristics related to the input physics labels. This behavior was enforced indirectly through the extended generator loss (\ref{eq:aux_loss}).
	By contrast, any other physically motivated observable evaluated on the generated showers was not constrained in the training. However, as illustrated in section \ref{subsec:observables}, many distributions of shower characterizing quantities computed on the generated showers match those computed from the real dataset well. 
    Moreover, we demonstrate that key correlations between calorimeter observables are obtained. For all reported benchmark scenarios, good shower qualities for $70~$GeV electron showers are obtained despite the fact that these were not part of the training set.

	Finally, this section is concluded with a report on the WGAN's computational time advantage over detailed simulation using GEANT4.

	\subsection{Visual inspection of generated showers}\label{subsec:displays}
	Figure \ref{fig:eventDisplayWGAN} shows two exemplary electron-induced showers with $20~$GeV and $90~$GeV energy labels generated using the WGAN approach. 
	
	This set of energy depositions is consistent with the physical intuition of how electron-induced cascades in this sampling calorimeter configuration should develop. First, it is noted that both pixel occupancies and pixel intensities scale with the incident electron energy. Second, the positions of the largest energy depositions move according to the input impact position labels. Finally, it is evident that the main activity of generated showers occurs in the central sampling compartments. In particular, the spread and the scale of energy depositions is maximal in intermediate layers, while only a few pixels are active in the first and last layers.

	Figure \ref{fig:AverageOccupancy} shows the average pixel occupancy with energy depositions above the 2 MIP threshold of $90~$GeV electron-induced showers simulated using GEANT4 compared to those generated by the WGAN. White spaces correspond to areas of the sensors which are activated above the threshold in less than $1\%$ of the events. The radial development of the pixel occupancy of WGAN-generated showers is similar to GEANT4 while the overall scale appears underestimated.

	\subsection{Label dependency}\label{subsec:labels}

	Three physics labels, namely the incident's electron energy $E$ and its impact position $P=(P_x, P_y)$, are input to the WGAN. Ideally, these labels should constrain the shower generation process. As described in section \ref{subsec:conditioning}, two constrainer networks are trained with real samples for this purpose and then reconstruct these labels based on the full shower information. For the training to be rated as successful, the reconstructed labels of the generated showers should correlate with the imposed physics labels. 
	Figure \ref{fig:dep_energy} shows the distribution of reconstructed energies for different energy labels. Their maxima correlate to the true labels. Furthermore, the energy spectra computed on WGAN generated and GEANT4 simulated showers exhibit a reasonable agreement.
	Figures \ref{fig:dep_posX} and \ref{fig:dep_posY} show the correlation for position labels. Here, the symbols indicate the mean reconstructed label in bins of the true label. 
	On average, a generated shower with a certain set of labels is reconstructed accordingly.
	Evidently, shower characteristics which the two constrainer networks are sensitive to are able to condition the generation process. Even the $70~$GeV electron cascades, which were not considered in the training, exhibit the same behavior.

	\begin{figure*}[t!]
		\captionsetup[subfigure]{aboveskip=-1pt,belowskip=-1pt}
		\begin{centering}
			\begin{subfigure}[b]{0.32\textwidth}
				\includegraphics[trim={0cm 0 0cm 0cm},clip,,width=\textwidth]{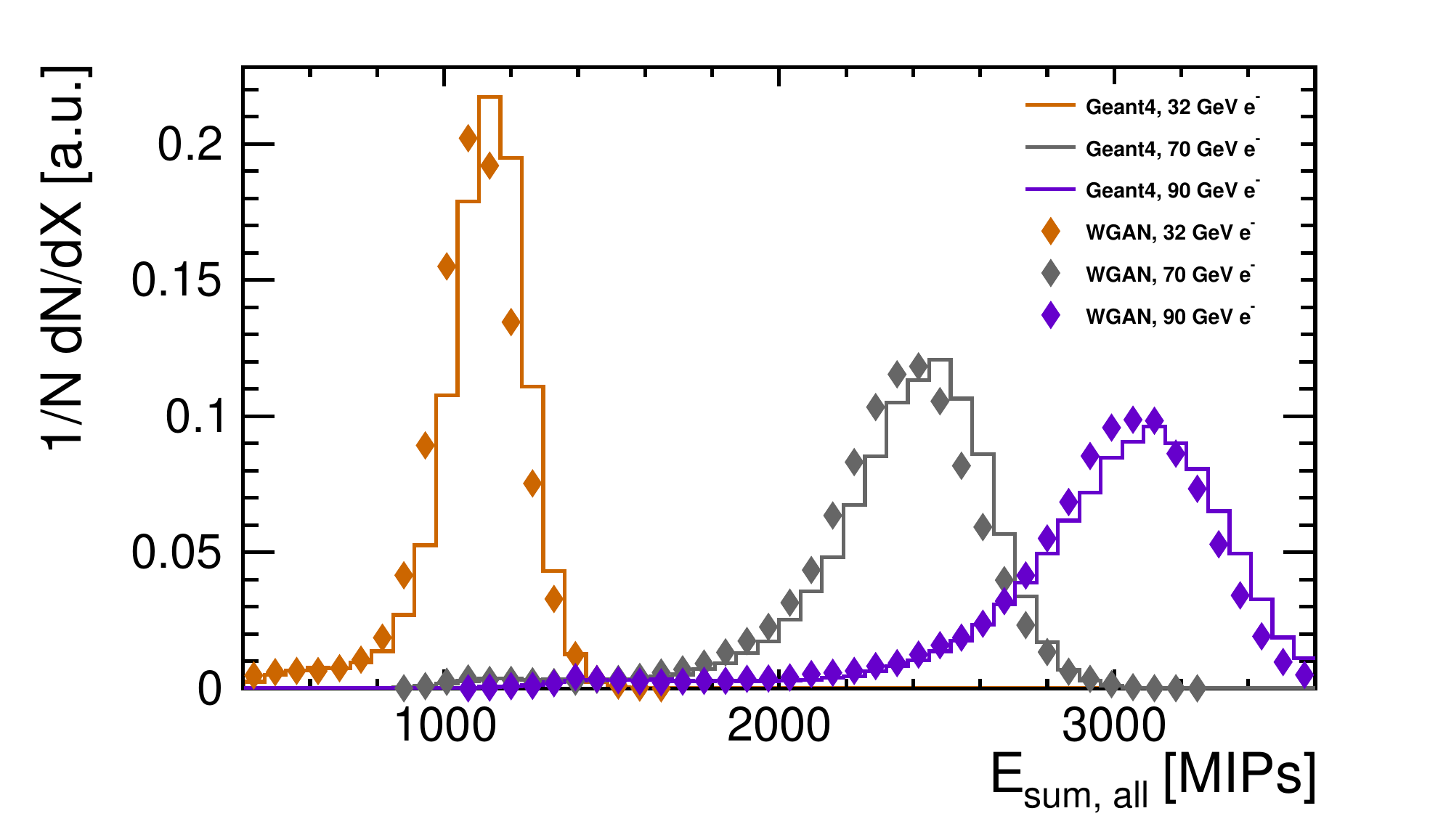}
				\subcaption{}
				\label{fig:Etot}
			\end{subfigure}
			\hfill
			\begin{subfigure}[b]{0.32\textwidth}
				\includegraphics[trim={0cm 0 0cm 0cm},clip,,width=\textwidth]{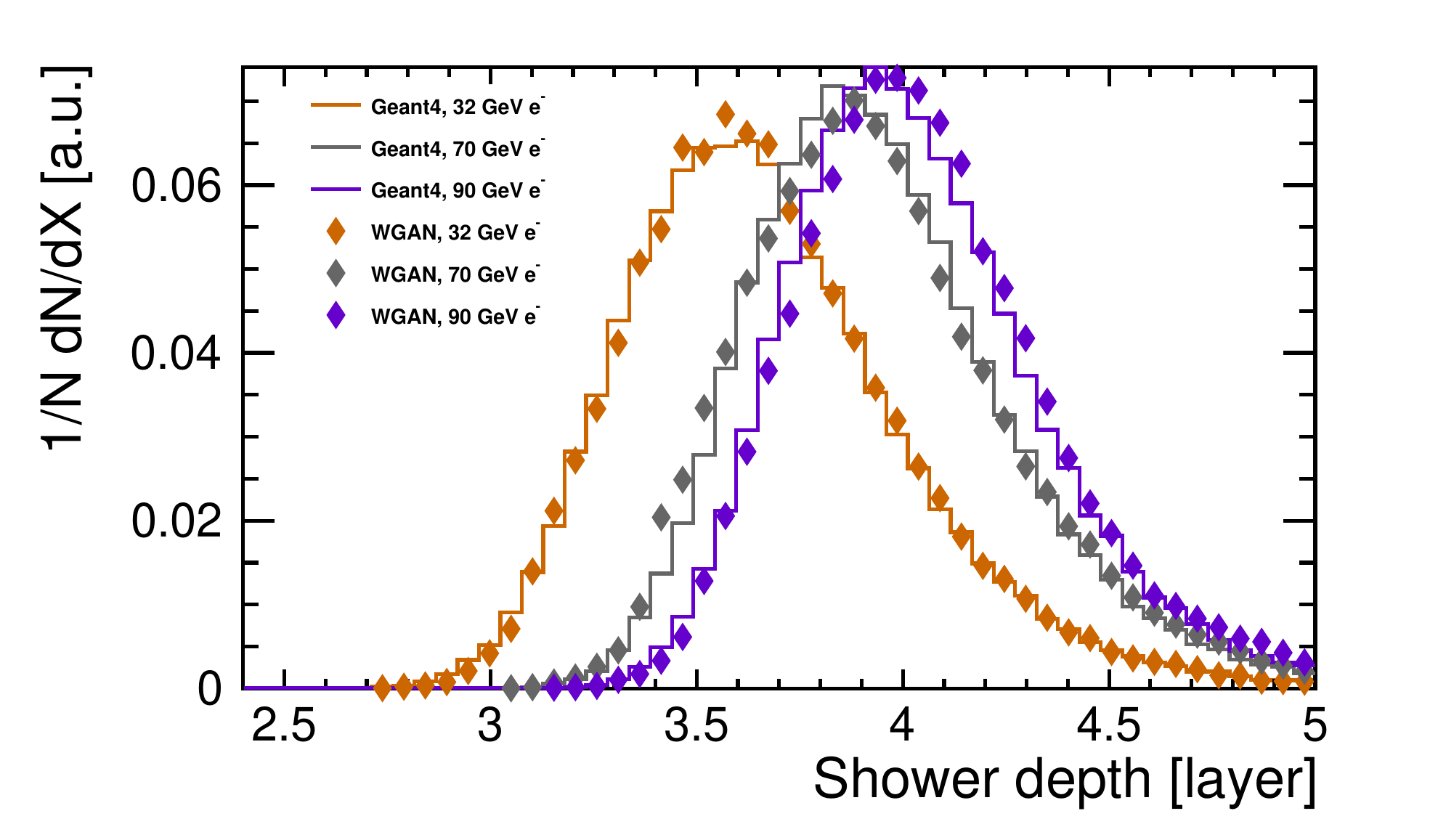}
				\subcaption{}
				\label{fig:X0}
			\end{subfigure}
			\hfill
			\begin{subfigure}[b]{0.32\textwidth}
				\includegraphics[trim={0cm 0 0cm 0cm},clip,,width=\textwidth]{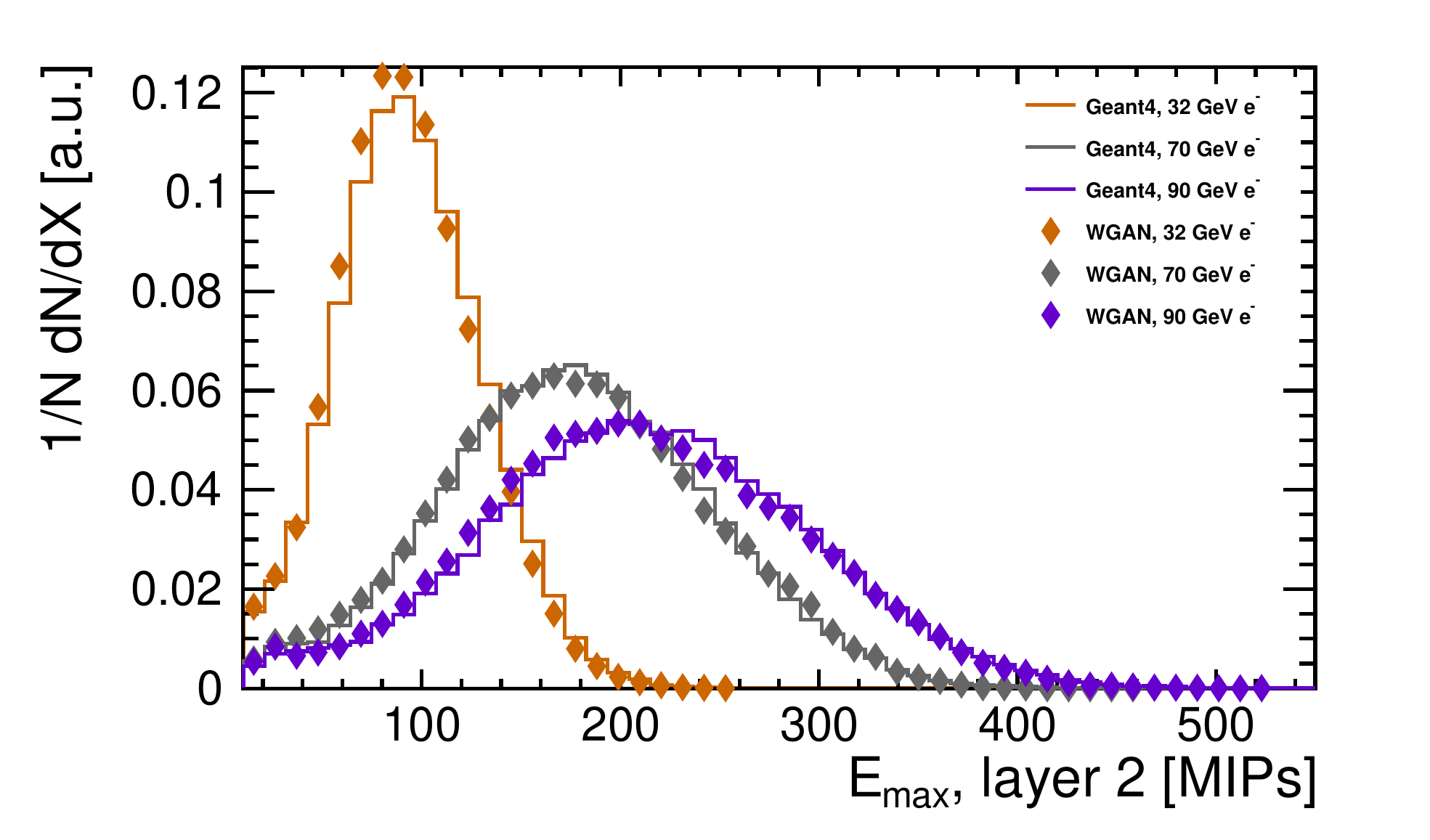}
				\subcaption{}
				\label{fig:Emax2}
			\end{subfigure}
			\hfill
			\begin{subfigure}[b]{0.32\textwidth}
				\includegraphics[trim={0cm 0 0cm 0cm},clip,,width=\textwidth]{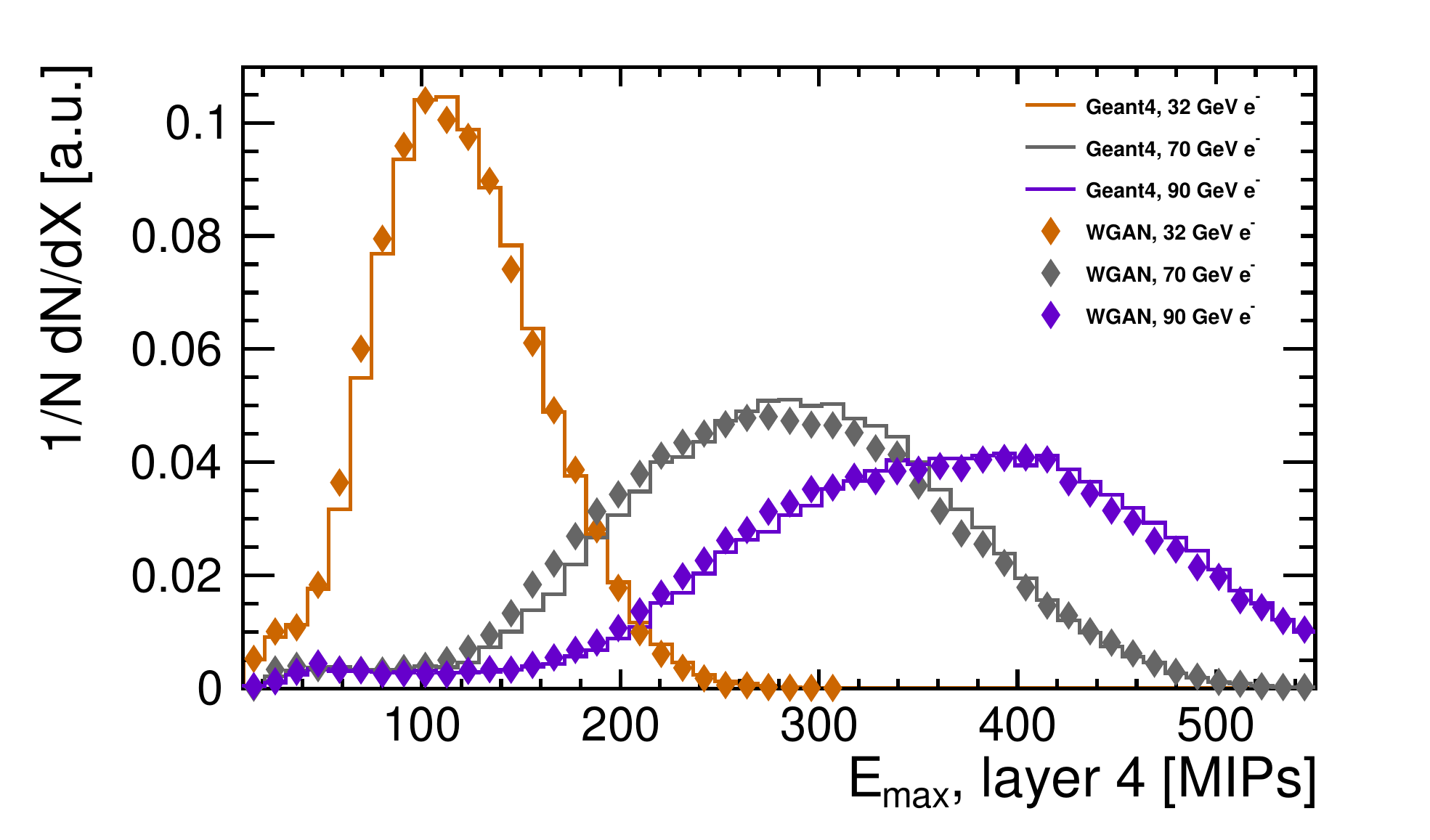}
				\subcaption{}
				\label{fig:Emax4}
			\end{subfigure}			
			\hfill
			\begin{subfigure}[b]{0.32\textwidth}
				\includegraphics[trim={0cm 0 0cm 0cm},clip,,width=\textwidth]{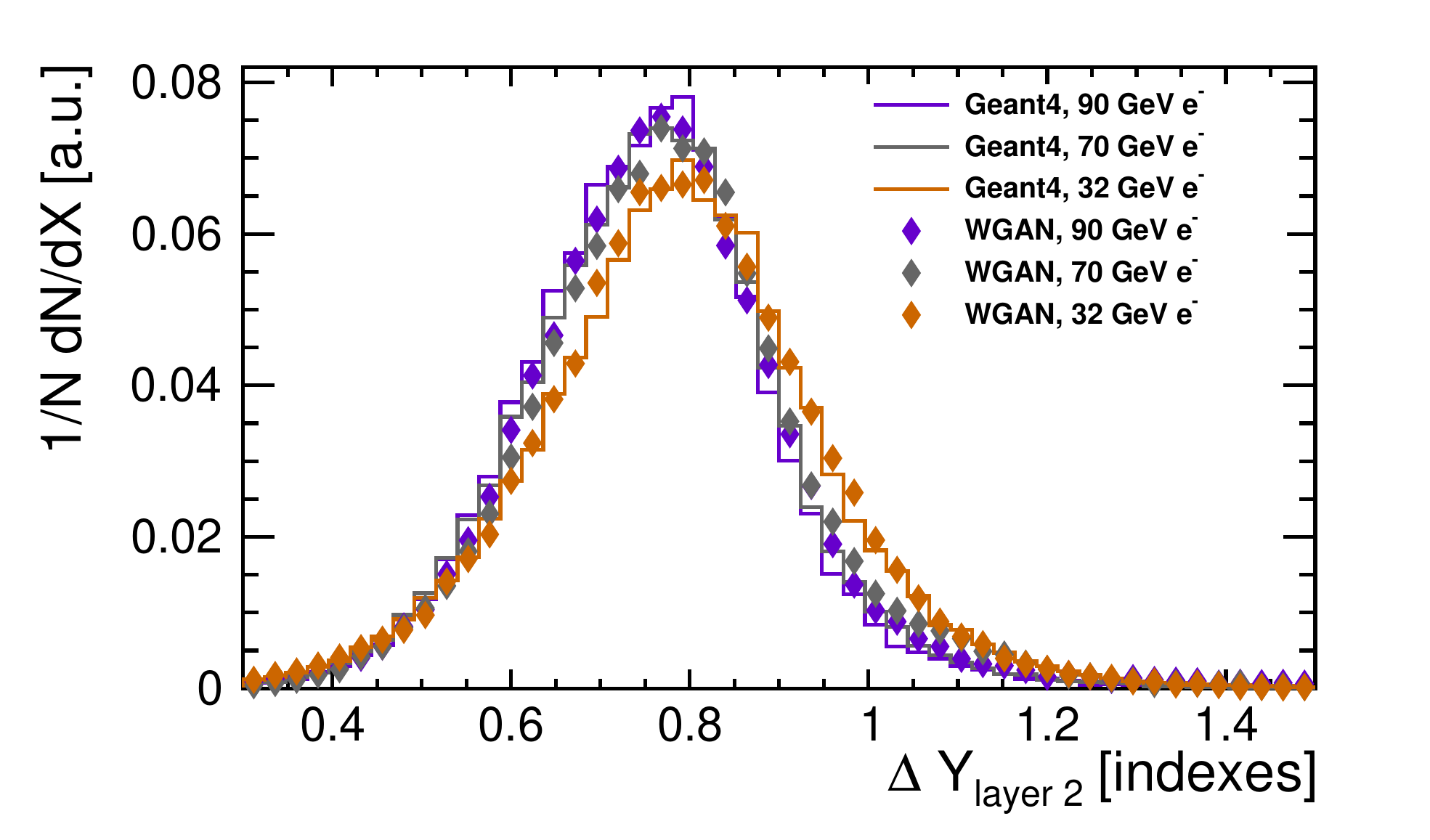}
				\subcaption{}
				\label{fig:DeltaY2}
			\end{subfigure}		
			\hfill
			\begin{subfigure}[b]{0.32\textwidth}
				\includegraphics[trim={0cm 0 0cm 0cm},clip,,width=\textwidth]{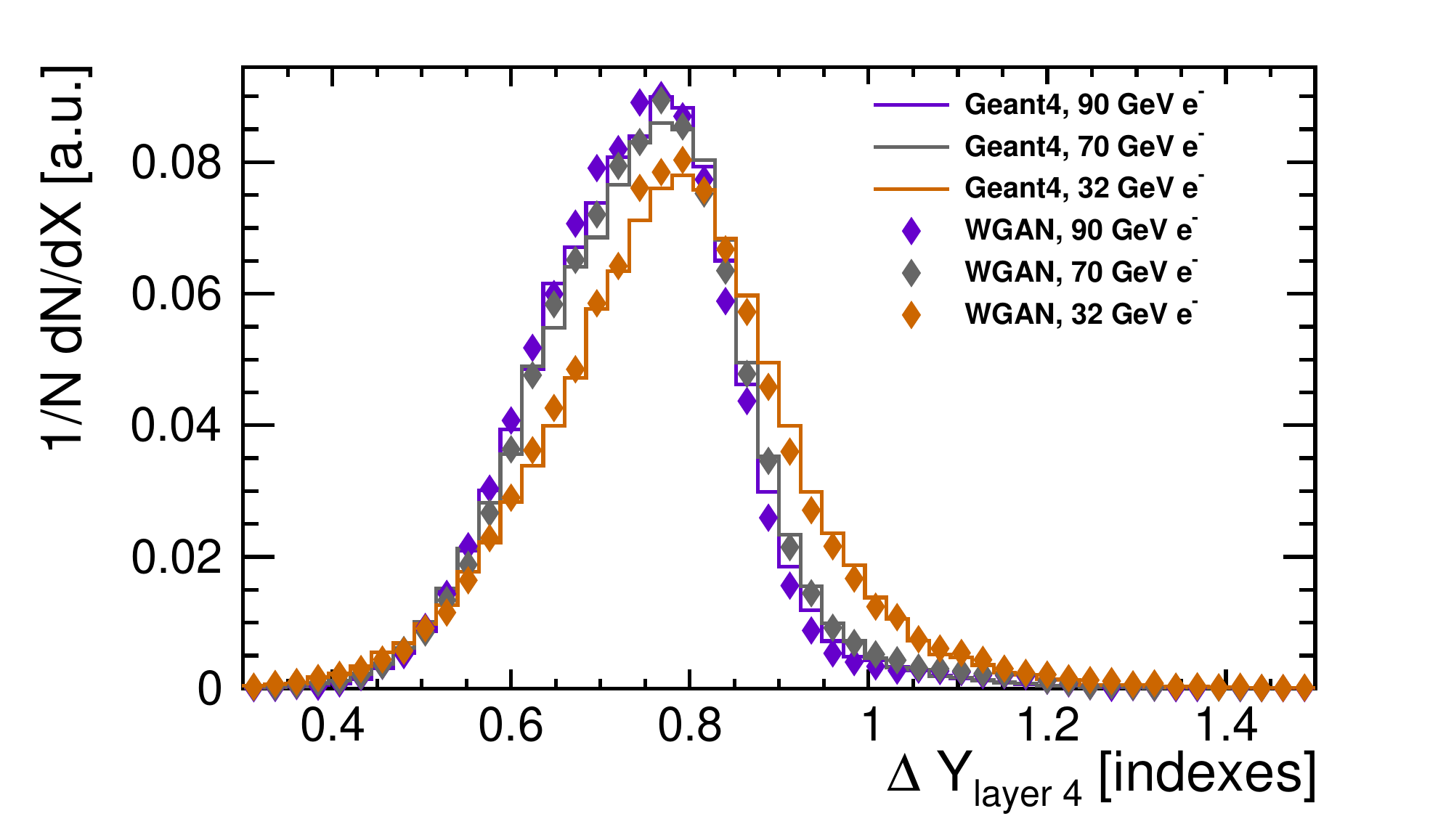}
				\subcaption{}
				\label{fig:DeltaY4}
			\end{subfigure}
			\caption{Comparison of calorimeter observables computed in generated showers (symbols) to those computed in fully simulated showers using GEANT4 (histograms). (a) Energy sum of all pixels, (b) energy-weighted shower depth, (c, d) the maximum pixel energy in layer 2, respectively 4, and (e, f) the transverse shower spread along the y-direction in layer 2, respectively 4. The $70~$GeV showers were not part of the training set.}
			\label{figure:OneDComparison}
		\end{centering}
	\end{figure*}

	\subsection{Calorimeter observables}\label{subsec:observables}
	In this section, typical calorimeter observables are presented which are computed both in GEANT4 simulations and WGAN-generated showers. 
	For better clarity, only the $32~$GeV and $90~$GeV as well as the additional $70~$GeV electron samples are shown. The agreement between real and generated cascades illustrated therein are representative for the entire dataset. Note that $70~$GeV showers were not part of the training set.
	
	\subsubsection{Distributions of calorimeter observables}\label{subsubsec:observablesCalo}
	Figure \ref{figure:OneDComparison} shows six sets of representative observables that characterize particle showers in sampling calorimeter configurations. The distributions are normalized to unity.
	
	A reasonable agreement between WGAN- and GEANT4-simulated showers is seen in Figure \ref{fig:Etot} for the total energy deposition summed over all pixels and  in Figure \ref{fig:X0} for the longitudinal shower depth. Also, the maximum pixel energy for each individual layer exhibits a good match with the full simulation (only layers 2 and 4 are shown in Figures \ref{fig:Emax2},  \ref{fig:Emax4}). \newline Furthermore, we compute the energy-weighted transverse spread in each layer. 
	\begin{equation}
	\Delta \text{Y}_{l}~=~\sum_{\text{pixel~}i}^{\text{layer } l} \big| y_i - \sum_{\text{pixel~}j}^{\text{layer }l} y_j\cdot \frac{E_j}{E_{\text{sum, }l}}\big| \cdot \frac{E_i}{E_{\text{sum, }l}}
	\label{eq:spread}
	\end{equation}
	$\Delta Y_{\text{layer~2}}$ and $\Delta Y_{\text{layer~4}}$ are shown here (\ref{fig:DeltaY2}, \ref{fig:DeltaY4}) by way of example. The computation for other layers $l$ and for the x coordinate is analogous. With the exception of the first layer at 2.8 $\text{X}_\text{0}$, the agreement therein is representative for all other layers and the x-coordinate. Thus the transverse shower shapes are well-modeled by the WGAN. 
\begin{figure}[t!]
	\captionsetup[subfigure]{aboveskip=-1pt,belowskip=-1pt}
	\begin{centering}
		\begin{subfigure}[b]{0.455\textwidth}
			\includegraphics[trim={0cm 0 0cm 0cm},clip,,width=\textwidth]{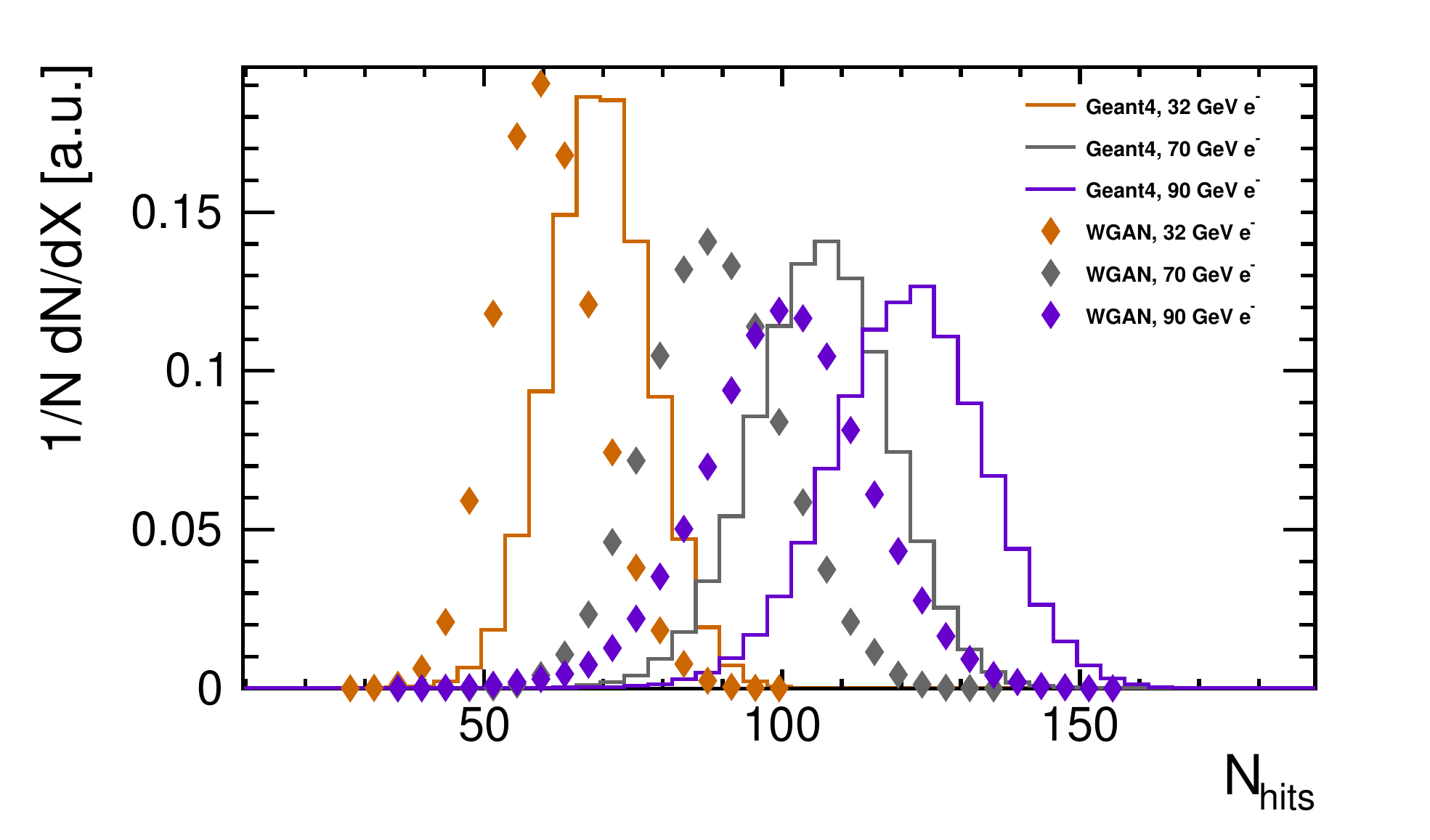}
			\subcaption{}
			\label{fig:Nhits}
		\end{subfigure}
		\hfill
		\begin{subfigure}[b]{0.455\textwidth}
			\includegraphics[trim={0cm 0 0cm 0cm},clip,,width=\textwidth]{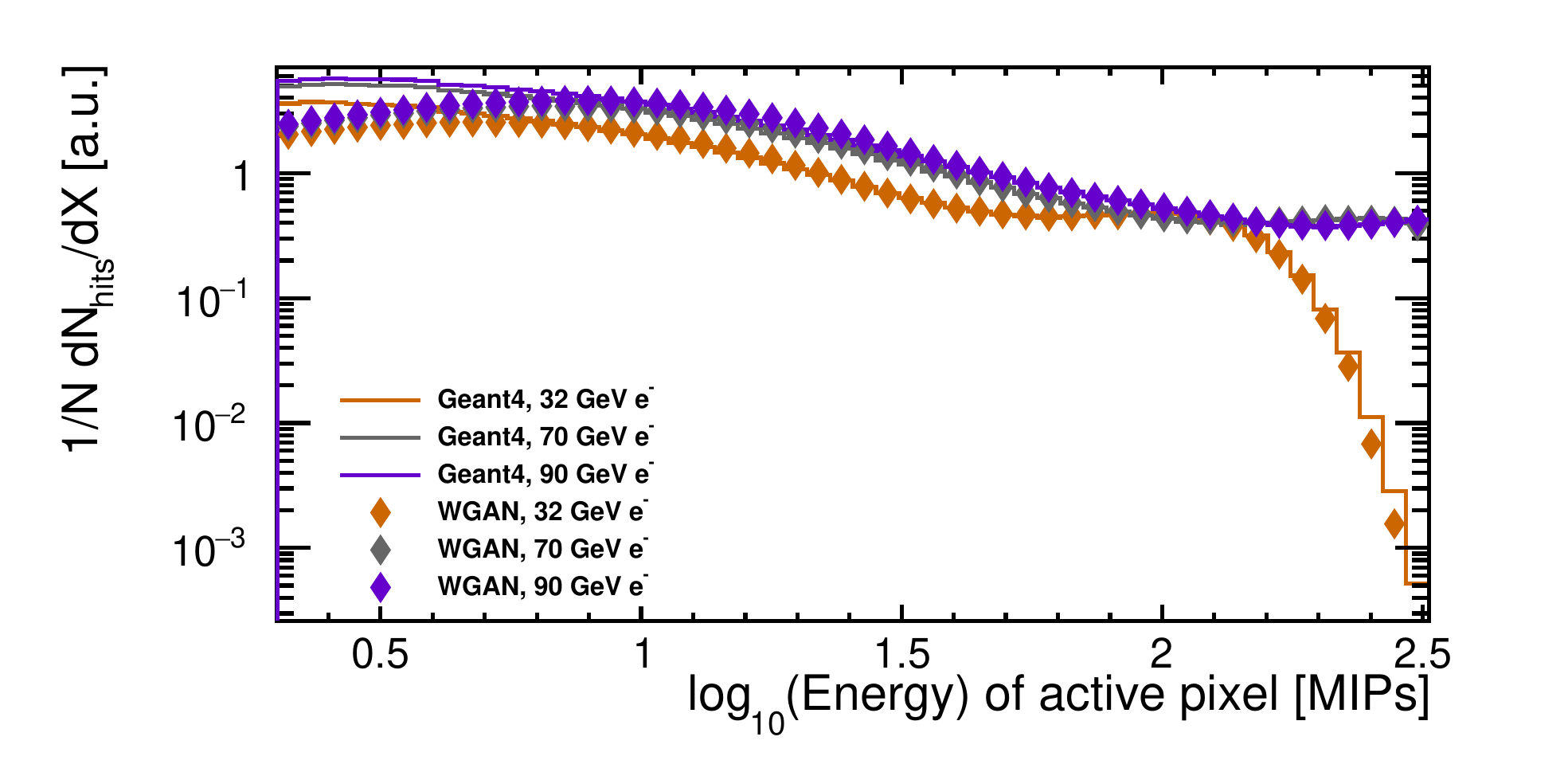}
			\subcaption{}
			\label{fig:Nspectrum}
		\end{subfigure}
		\caption{(a) Distributions of the number of pixels with energy depositions above 2 MIPs equivalents for several energies. (b) Single pixel energy spectra show reasonable agreement with the simulation for energy densities above $\approx 10~$MIPs per pixel. Note that $70~$GeV showers were not part of the training set.}
		\label{figure:NhitsComparison}
	\end{centering}
\end{figure}    

\begin{figure*}[t!]
	\captionsetup[subfigure]{aboveskip=-1pt,belowskip=-1pt}
	\begin{centering}
		\begin{subfigure}[b]{0.32\textwidth}
			\includegraphics[trim={0cm 0 0cm 0cm},clip,,width=\textwidth]{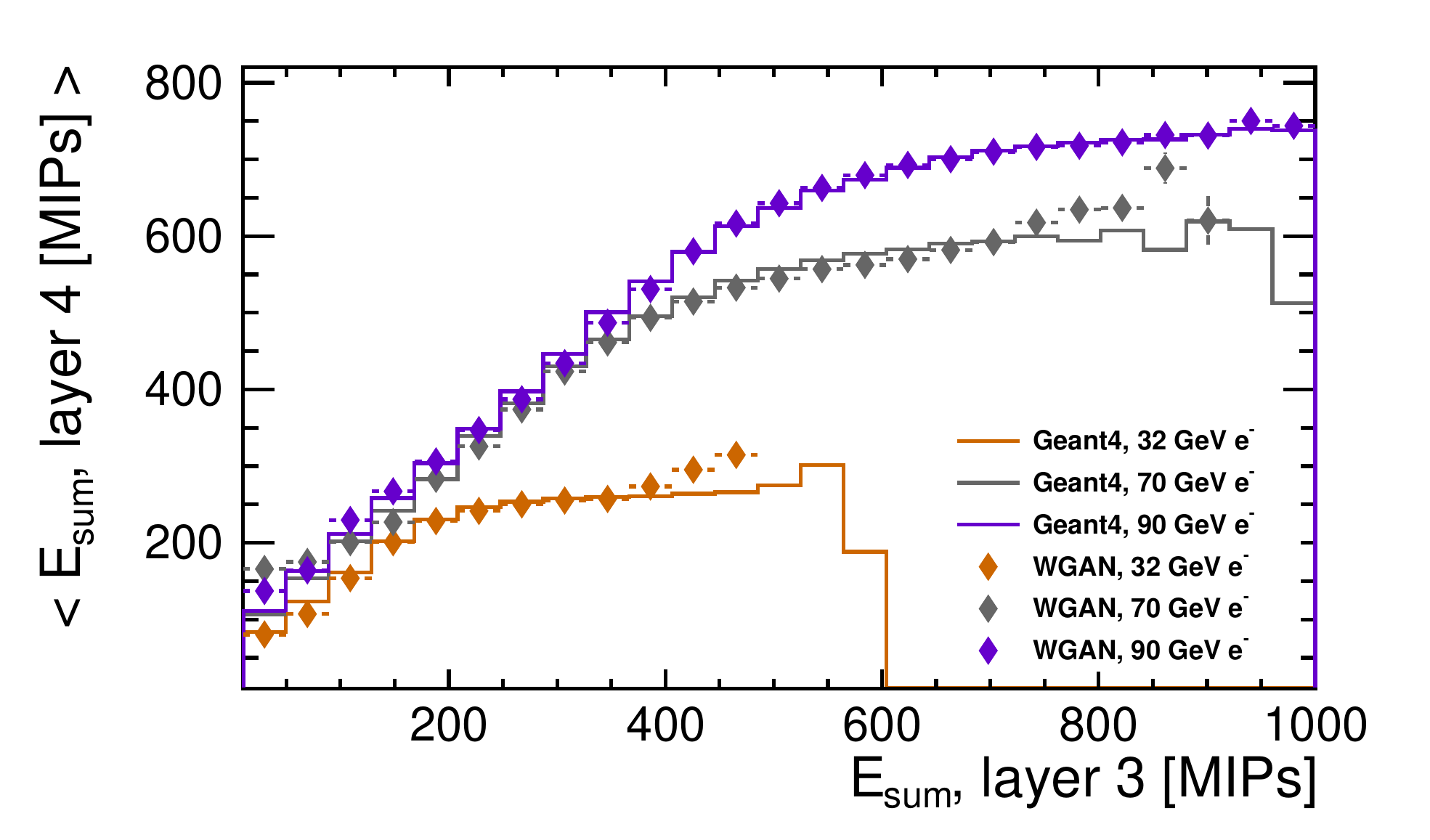}
			\subcaption{}
			\label{fig:Etot_3_4}
		\end{subfigure}
		\hfill
		\begin{subfigure}[b]{0.32\textwidth}
			\includegraphics[trim={0cm 0 0cm 0cm},clip,,width=\textwidth]{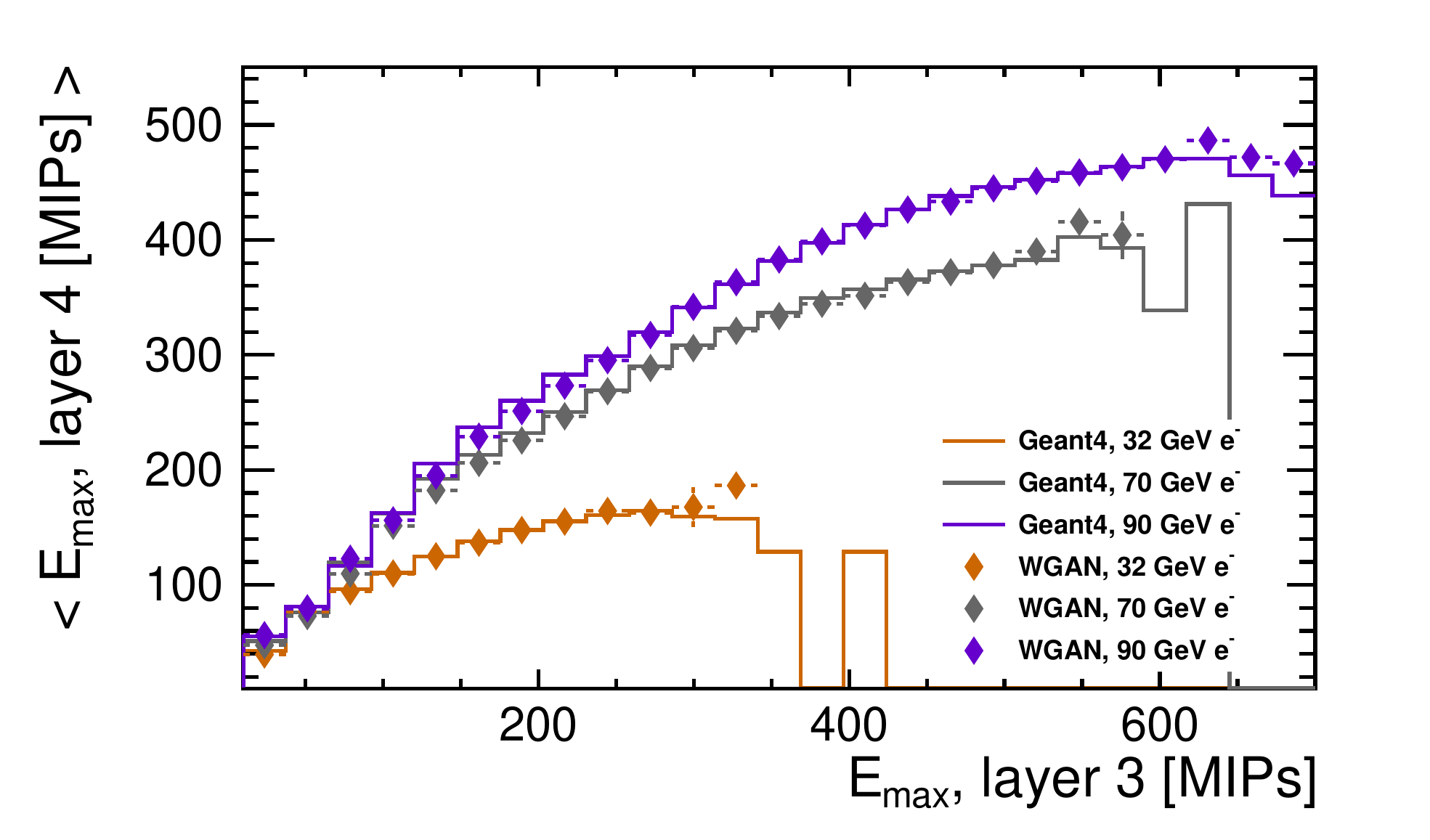}
			\subcaption{}
			\label{fig:Emax_3_4}
		\end{subfigure}	
		\hfill
		\begin{subfigure}[b]{0.32\textwidth}
			\includegraphics[trim={0cm 0 0cm 0cm},clip,,width=\textwidth]{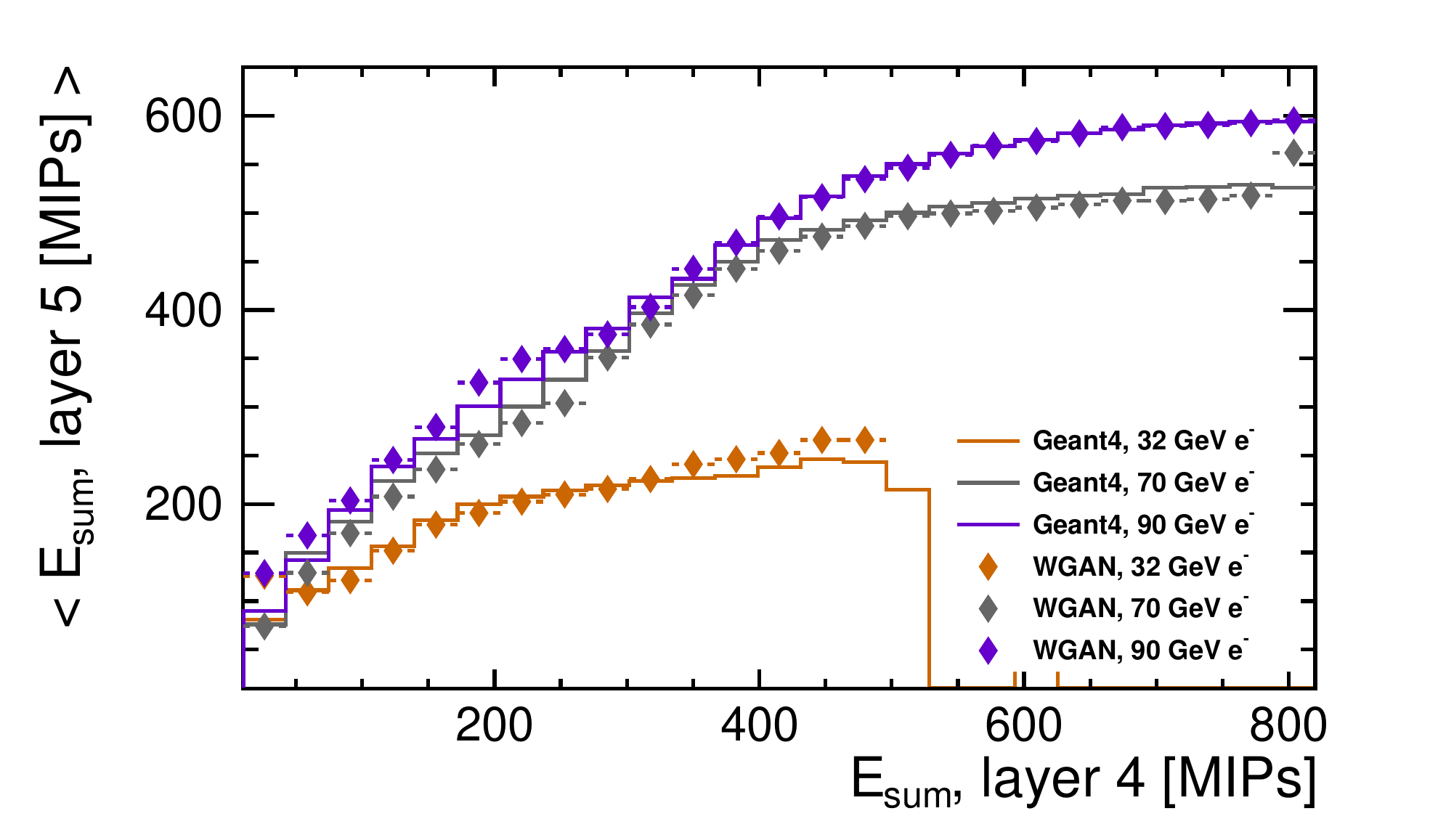}
			\subcaption{}
			\label{fig:Etot_4_5}
		\end{subfigure}
		\begin{subfigure}[b]{0.32\textwidth}
			\includegraphics[trim={0cm 0 0cm 0cm},clip,,width=\textwidth]{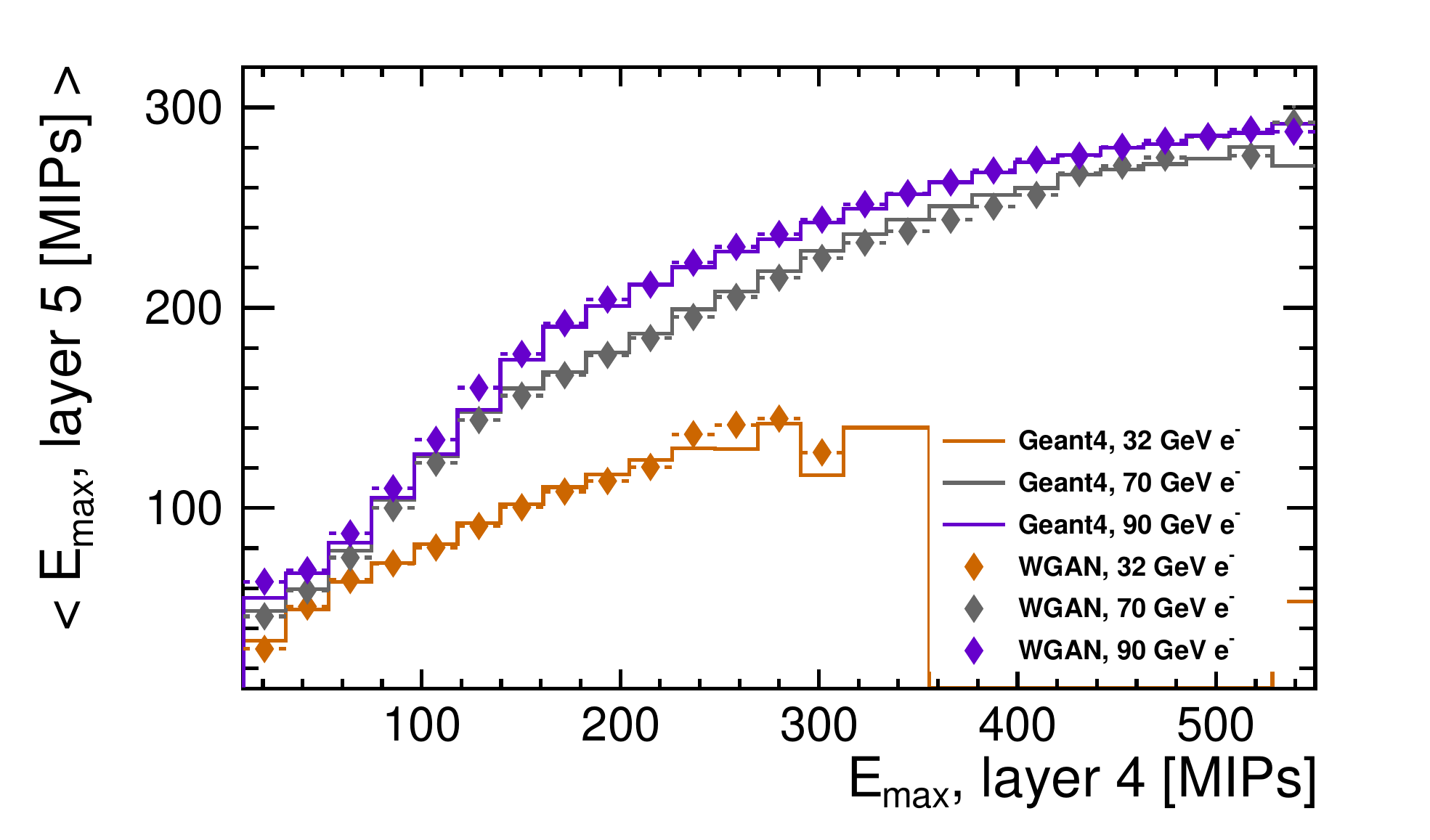}
			\subcaption{}
			\label{fig:Emax_4_5}
		\end{subfigure}
		\hfill		
		\begin{subfigure}[b]{0.32\textwidth}
			\includegraphics[trim={0cm 0 0cm 0cm},clip,,width=\textwidth]{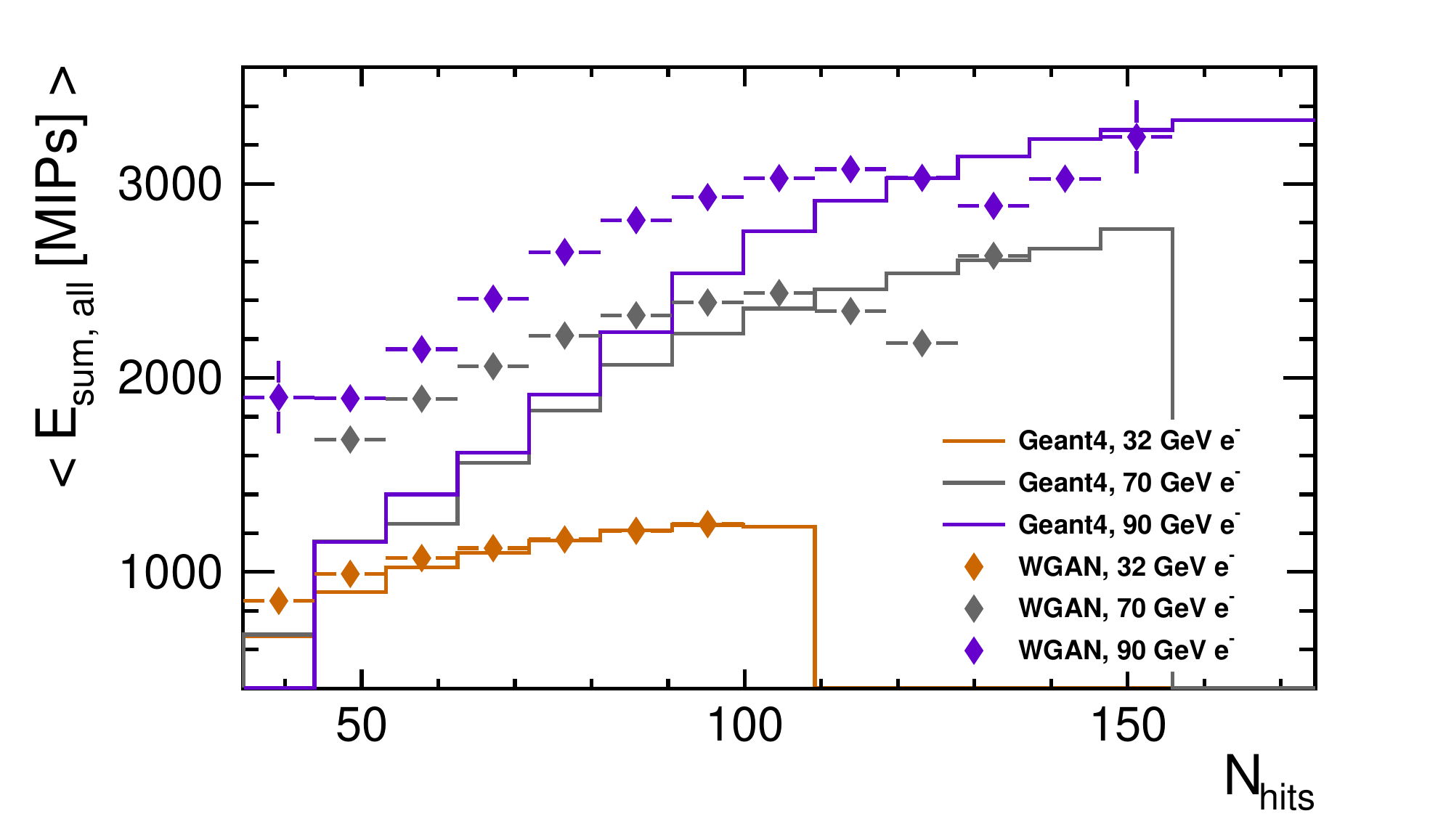}
			\subcaption{}
			\label{fig:Etot_Ntot}
		\end{subfigure}
		\hfill
		\begin{subfigure}[b]{0.32\textwidth}
			\includegraphics[trim={0cm 0 0cm 0cm},clip,,width=\textwidth]{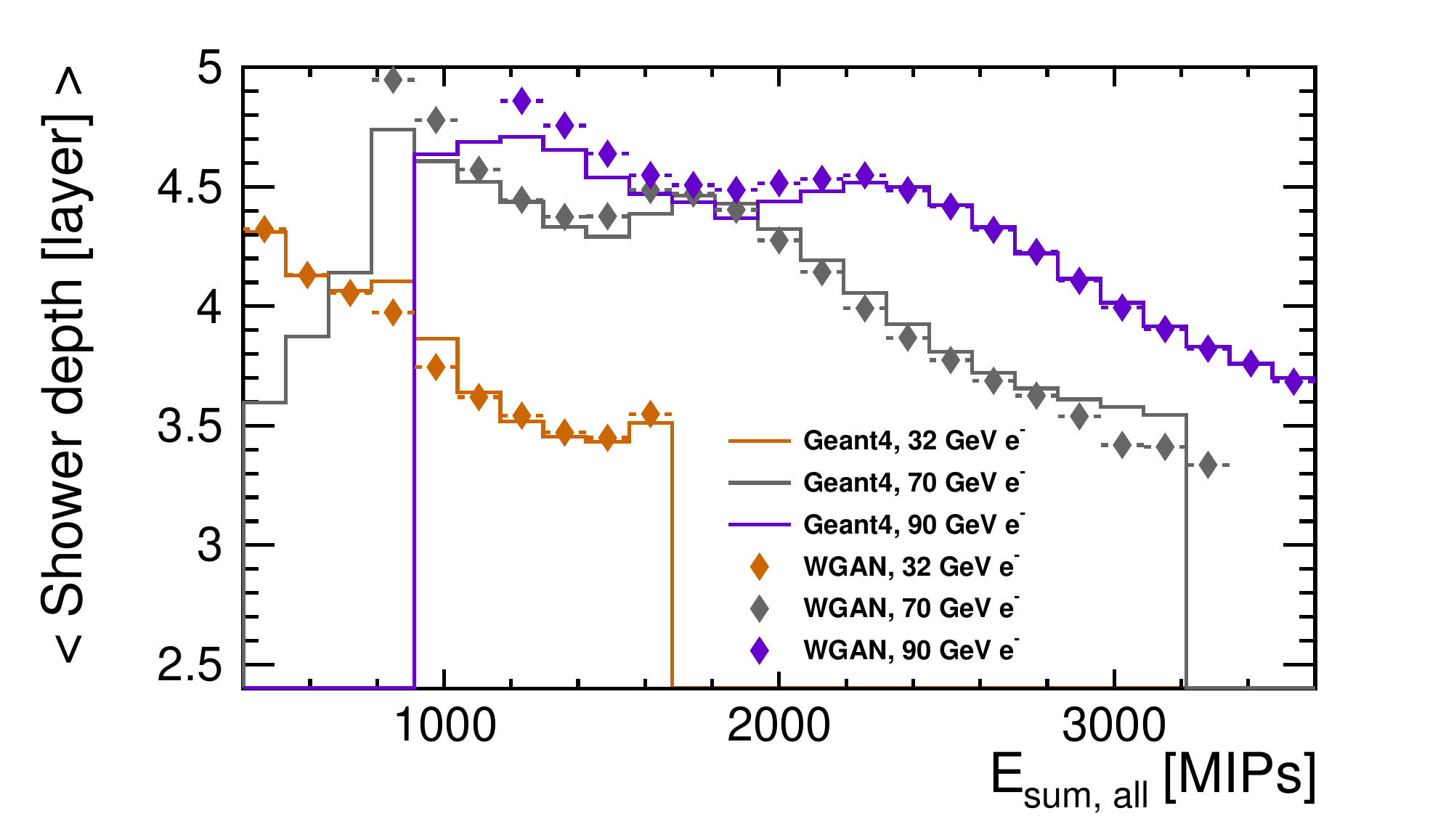}
			\subcaption{}
			\label{fig:Etot_X0}
		\end{subfigure}
		\caption{Comparison of calorimeter observable correlations evaluated on generated and GEANT4-simulated showers. (a) Energy sum, respectively (b) maximum pixel energy, in layer 4 vs. total energy deposit (maximum pixel energy) in layer 3. (c) Energy sum, respectively (d) maximum pixel energy, in layer 5 vs. total energy deposit (maximum pixel energy) in layer 4. (e) Total energy deposit plotted against number of hits. (f) Shower depth vs. total energy deposit. Note that $70~$GeV showers were not part of the training set.}
		\label{fig:correlations}
	\end{centering}
\end{figure*}

    In Figure \ref{fig:Nspectrum} we investigate the energy spectrum of active pixels. The region of low energy densities, i.e. pixels with depositions below $\approx 10~$MIPs, is underrepresented by the WGAN with respect to GEANT4 causing a mismatch in the number of active pixels (energy $\geq 2~$MIPs) in Figure \ref{fig:Nhits} and ultimately also resulting in the underestimate of their energy sum (Figure \ref{fig:Etot}). Similar mismodeling of sparsity-describing quantities has also been reported in the work based on traditional GANs \cite{Hooberman:2017nips,Paganini:2017dwg}. 
	It should be noted that the analysis of such fast simulated showers could always be limited to the well-described range by restricting the analysis to pixels with energies above 10 MIP equivalents. In this calorimeter setup, the rejected part of the spectrum contributes only 10\% to the total signal. This could be corrected by scale factors.

	Following this principle, we conducted a supplementary performance benchmark. Example graphics are shown in the appendix (Figures \ref{fig:AppendixAverageOccupancy}, \ref{fig:Appendixcorrelations}). The agreement between the WGAN and GEANT4 in this regime is improved also for all other observables.

	\subsubsection{Correlations}\label{subsubsec:correlation} 		         	  
	For energy reconstruction or particle identification in modern particle detection systems, multiple calorimeter observables are needed simultaneously  (e.g. \cite{ATLAS:2005NIMPR}). Typical approaches exploit correlations of shower characteristics. Consequently, a crucial quality measure for a simulation tool is the assessment of the pairwise correlations of reconstructed physics observables.

	In this section, we focus on four examples of such correlations. 
	First, both the summed energy and the maximum pixel energy in a fixed layer is expected to correlate with the sum (maximum) in the previous layer. Fig. \ref{fig:Etot_3_4} - \ref{fig:Emax_4_5} visualize this trend for layers 3 and 4, respectively 4 and 5, by way of example. It must be noted that the WGAN-generated showers exhibit good agreement with the GEANT4 showers. 
	
	Second, a greater number of active pixels should correspond to higher values in the sum of energy depositions, which is illustrated in Figure \ref{fig:Etot_Ntot}. Some discrepancy is to be expected owing to the underestimation of the low energy spectrum by the WGAN. While the positive correlation is also obtained for the WGAN samples, an agreement is not reached. 
	
	Ultimately, the sampling in this specific calorimeter configuration is not uniform as the setup comprises non-equidistant sampling layers. As a consequence, showers with large energy depositions at a fixed incident electron energy should relate to lower shower depths. Showers whose center of gravity is located deeper in the calorimeter deposit higher fractions of their energy in the larger amount of passive material between layer 5 and 7, hence resulting in lower values of energy sums in the sensitive layers. 
	Also for this example, a good match between real and WGAN-generated showers is displayed in Figure \ref{fig:Etot_X0}. 

	\begin{table*}[t!]
	\caption{Computational time required for the generation of one $20~$GeV, respectively $90~$GeV, electron-induced cascade through evaluation of the WGAN using different hardware setups and enhancement with respect to a full simulation using GEANT4. Using the WGAN approach, a speed-up of more than three orders of magnitude has been achieved.}
	\centering
	\begin{tabular}{cc|cc|cc}
		Method & Hardware & $20~$GeV $\text{e}^{-}$ & \textbf{Speed-up} & $90~$GeV $\text{e}^{-}$ & \textbf{Speed-up} \\
		\toprule
		GEANT4 & any CPU & $\orderof\left(500~\text{ms} \right)$ & - & $\orderof\left(2000~\text{ms} \right)$ & - \\
		WGAN  & Intel\textsuperscript{\textcopyright} Xeon\textsuperscript{\textcopyright} CPU E5-1620 & 52 ms & \textbf{x10} & 52 ms  & \textbf{x40} \\				
		WGAN & NVIDIA\textsuperscript{\textcopyright} Quadro\textsuperscript{\textcopyright} K2000 GPU  & 3.6 ms & \textbf{x140} & 3.6 ms & \textbf{x560} \\				
		WGAN & NVIDIA\textsuperscript{\textcopyright} GTX\textsuperscript{\texttrademark} 1080 GPU  & 0.3 ms & \textbf{x1660} & 0.3 ms & \textbf{x6660} \\				
	\end{tabular}
	\label{table:computing}
\end{table*}

	In summary, typical calorimeter observables computed in fast WGAN-generated showers correspond well to those simulated using GEANT4, not only in terms of their spectra but also in their pairwise correlations. Only the number of low-energy depositions is too small. A detailed simulation typically requires extensive tuning of its model parameters using expert knowledge in particle interactions with matter to achieve a similar level of agreement to real data. By contrast, no dedicated knowledge on how particles interact with the material had to be input into the generative model here.

	\subsection{Computational speed-up}\label{subsec:computing}
	We observe that fast simulation of electromagnetic showers using this WGAN architecture is up to three orders of magnitude faster than full simulations. By contrast, expert engineered, parameterized fast simulation provides CPU gains of 10-100 with respect to GEANT4 depending on the particle energy, see e.g. \cite{CMS:FastSim}. In Table \ref{table:computing}, we provide numbers on the computation time advantage of our WGAN compared to GEANT4 using three different hardware architectures. As expected, graphics processing units (GPUs) are the preferred hardware unit since the WGAN's internal set of linear computational operations is more efficiently parallelized and runs faster than in CPUs. Furthermore, it is remarkable that the time for evaluation of the WGAN is independent of the incident electron energy while it scales significantly in simulations using GEANT4. \newline

\section{Conclusion}\label{sec:Conclusion}
    
In this paper, we presented a method for generating electromagnetic showers using a Wasserstein GAN (WGAN). As a concrete example, we followed a prototype setup for a future high-granularity calorimeter with seven sensitive layers featuring about one thousand readout pixels placed in an electron beam. As a generative model, we constructed a system of two adversarial networks, one being the generator and the other ensuring high-quality showers. The latter is called the critic network and was active only during the training phase where the probability distribution encoded in reference data was transferred to the generator. The critic network is designed to provide an
approximation of the Wasserstein distance between generated and reference data. Using this concept, the training process was found to be well under control. Furthermore, two constrainer networks were included within the elaborate multi-network architecture. These constrainer networks ensure the desired dependencies of the generated showers with respect to the primary electron energies and impact locations on the calorimeter. They were also trained successfully within the elaborate multi-network architecture.

When benchmarking the WGAN-generated showers, visual inspections of single showers reveal the typical shower properties with high fluctuations and sparse energy depositions. The average pixel occupancy in the sensitive layers of WGAN-generated showers and of GEANT4 showers appears to be very similar. 

Various observables typically inspected for calorimeter showers exhibit good agreement with WGAN-generated and GEANT4 showers. For example, the shapes of the longitudinal shower depths are well reproduced not only for the different primary electron energies which the network had been trained for but also for an intermediate electron energy which the network had not encountered before. Also, the total energy depositions in each sensitive layer and the maximum energy in a sensor are described in detail. Furthermore, the transverse shower shapes appear as expected. Only the spectrum at low-energy densities is underestimated by the WGAN. Hence, the total number of sensor pixels with energies above a threshold energy equivalent to two minimum ionizing particles (MIPs) was found to be reduced by about $15\%$.

In addition, we analyzed correlations between energy depositions in the sensitive layers. Also here we established a strong agreement between the energies of neighboring layers for the different primary electron energies. The longitudinal shower depth decreases with increasing energy sums of all layers as expected for this specific sampling configuration. We also correlate the total energy sum with the total number of sensor pixels with energies above the 2 MIPs threshold. 

While this setup has only been studied in the context of isolated showers, there are ideas for its application to full collision events. In particular, it is straight forward to extend the set of labels limiting the WGAN simulation to a quantitative measure of background activity around a given shower. Alternatively, energy deposits of independent but simultaneously occurring particles in a sensor could be superimposed if electronic and saturation effects can be neglected in the underlying detection technology.

In the work presented here it was shown that a WGAN can be successfully used to simulate isolated electromagnetic showers in a realistic setup of a multi-layer sampling calorimeter. The computational speed-up compared to traditional sequential simulations amounts to several orders of magnitude. At the same time, in most aspects, the quality of these ultra-fast shower simulations with the WGAN reaches the level of showers generated with the GEANT4 program.

\newpage
\section*{Acknowledgments}

For valuable discussions and comments on the manuscript we wish to thank Lucie Linssen, Eva Sicking and Florian Pitters from the EP-LCD group at CERN, and Yannik Rath from the Aachen group. We gratefully acknowledge permission to apply the geometry files provided by the CMS HGCAL group for simulating data needed for this study.
This work is supported by the Ministry of Innovation, Science and Research of the State of North Rhine-Westphalia, and the Federal Ministry of Education and Research (BMBF). Thorben Quast gratefully acknowledges the grant of the Wolfgang Gentner scholarship.

\newpage
\eject
\appendix
\onecolumn
\section{Appendix}\label{sec:Appendix}
    
\begin{table}[!h]
\centering
\caption{Critic network as used in the adversarial framework.}
\label{table:critic}
 \begin{tabular}{l l l l l l}
 \toprule
 Operation & Kernel & Features & Padding & Normalization & Activation \\ \midrule
 \multicolumn{6}{l}{Critic input $12\times15\times7$ + $3$}\\ \midrule
 Linear & N/A & 10 nodes & N/A & $\times$ & leaky ReLU\\
 Linear & N/A & 180 nodes & N/A & $\times$ & leaky ReLU\\
 Reshape to $12\times 15 \times 1$ & & & & & \\ \midrule
 \multicolumn{6}{l}{Concatenation with input to $12\times 15 \times 8$}\\ \bottomrule
 Convolution 2D & $5\times$5 & 256 maps & same & $\times$ & leaky ReLU\\
 Convolution 2D & $3\times$3 & 128 maps & same & layer norm & leaky ReLU\\
 Convolution 2D & $3\times$3 & 64 maps & same & layer norm & leaky ReLU\\
 Convolution 2D & $3\times$3 & 32 maps & same & layer norm & leaky ReLU\\
 Convolution 2D & $3\times$3 & 16 maps & same & layer norm & leaky ReLU\\
 Linear & N/A & 10 nodes & N/A & $\times$ & leaky ReLU\\
 Linear & N/A & 1 node& N/A & $\times$ & $\times$\\ \midrule
 \multicolumn{6}{l}{Critic output $1$}\\ \bottomrule
\end{tabular}
\end{table}

\begin{table}[!h]
\centering
\caption{Generator network as used in the framework to generate electromagnetic calorimeter showers.}
\label{table:generator}
 \begin{tabular}{l l l l l l l}
 \toprule
 Operation & Kernel & Features & stride & Padding & Normalization & Activation \\ \midrule
 \multicolumn{6}{l}{Generator input $10+3$}\\ \midrule
 Tower $\times$7 & & & & & & \\
 Linear & N/A & 10 nodes & N/A & N/A & $\times$ & leaky ReLU \\
 Linear & N/A & 192 nodes & N/A & N/A & $\times$ & leaky ReLU \\
 Reshape to $3\times 4 \times 16$ & & & & & & \\
 Transposed convolution 2D & $3\times$3 & 16 maps & $2\times$2 & same & batch norm & leaky ReLU\\
 Transposed convolution 2D & $3\times$3 & 32 maps & $2\times$2 & same & batch norm & leaky ReLU\\
 Transposed convolution 2D & $3\times$3 & 64 maps & $2\times$2 & same & batch norm & leaky ReLU\\
 Convolution 2D & $5\times$9 & 1 maps & $1\times$1 & same & batch norm & leaky ReLU\\ \midrule
 Concatenation of 7 towers  & & & & & & \\
 Convolution 2D & $3\times$3 & 64 maps & $1\times$1 & same & batch norm & leaky ReLU\\
 Convolution 2D & $5\times$6 & 128 maps & $1\times$1 & valid & batch norm & leaky ReLU\\
 Convolution 2D & $3\times$3 & 14 maps & $1\times$1 & same & batch norm & leaky ReLU\\
 Locally connected 2D & $3\times$3 & 7 maps & $1\times$1 & same & $\times$ & ReLU\\ \midrule
 \multicolumn{6}{l}{Generator output $12\times 15 \times 7$}\\ \bottomrule
\end{tabular}
\end{table}

\begin{table}[!h]
\centering
\caption{Constrainer network as used in the framework for energy (position) regression.}
\label{table:constrainer}
 \begin{tabular}{l l l l l l}
 \toprule
 Operation & Kernel & Features & Padding & Normalization & Activation \\ \midrule
 \multicolumn{6}{l}{Constrainer input $12\times15\times7\times1$}\\ \midrule
 Convolution 3D & $3\times$3$\times$3 & 1 map & same & batch norm & leaky ReLU\\
 Convolution 3D & $3\times$3$\times$2 & 16 maps & same & batch norm & leaky ReLU\\
 Convolution 3D & $3\times$3$\times$2 & 16 maps & same & batch norm & leaky ReLU\\
 Convolution 3D & $3\times$3$\times$2 & 32 maps & same & batch norm & leaky ReLU\\
 Convolution 3D & $3\times$3$\times$2 & 32 maps & same & batch norm & leaky ReLU\\
 Convolution 3D & $3\times$5$\times$2 & 64 maps & same & $\times$ & leaky ReLU\\
 Linear & N/A & 1(2) node(s) & N/A & $\times$ & $\times$ \\ \midrule
 \multicolumn{6}{l}{Constrainer output $1$($2$)}\\ \bottomrule
\end{tabular}
\end{table}

\newpage

\begin{figure}[H]
\centering 
\includegraphics[width=0.7\textwidth]{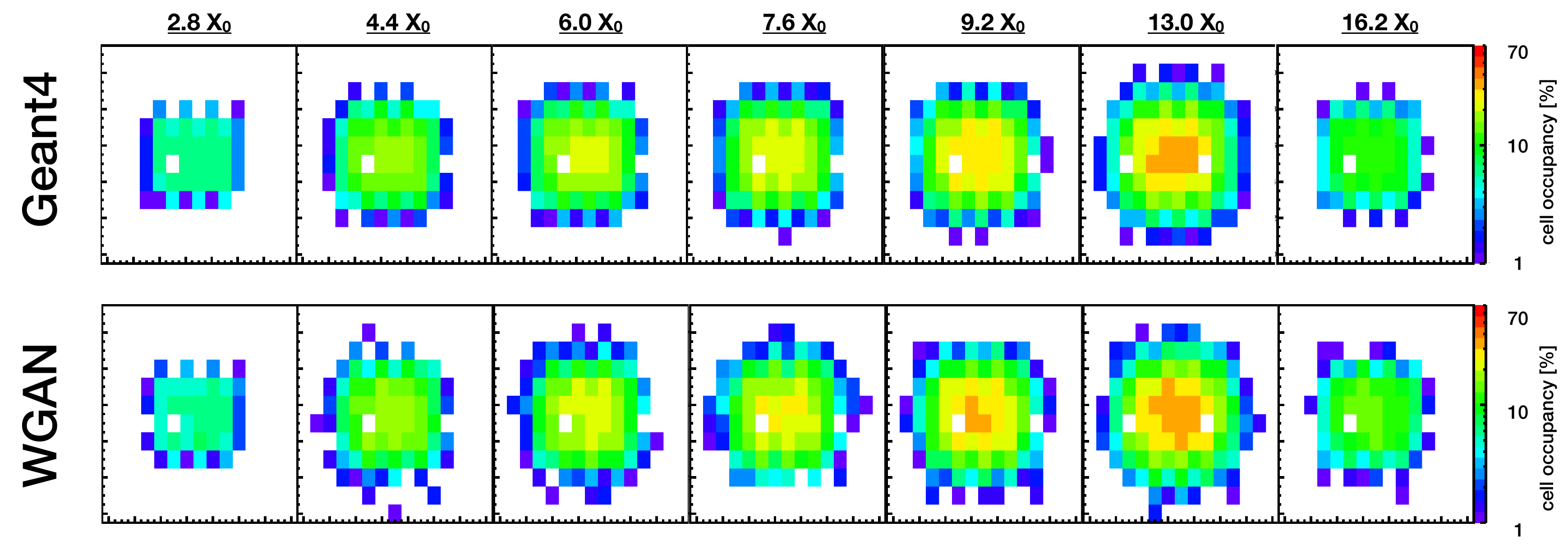}
\caption{ Analogous to Figure \ref{fig:AverageOccupancy}. The noise threshold is set to 10 MIPs instead of 2 MIPs. }
\label{fig:AppendixAverageOccupancy}	
\end{figure}

\begin{figure}[H]
	\captionsetup[subfigure]{aboveskip=-1pt,belowskip=-1pt}
	\begin{centering}
		\begin{subfigure}[b]{0.32\textwidth}
			\includegraphics[trim={0cm 0 0cm 0cm},clip,,width=\textwidth]{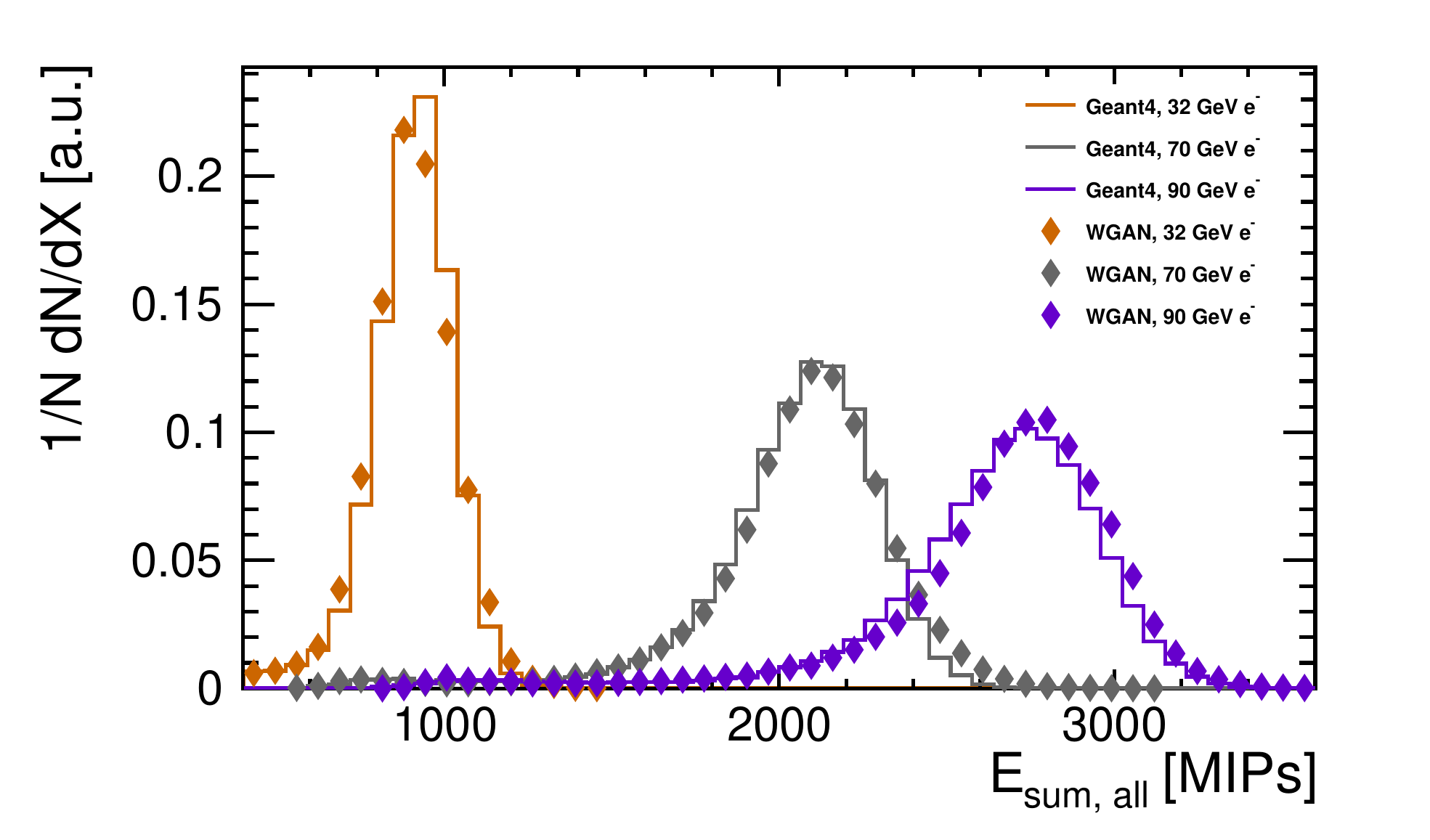}
			\subcaption{}
		\end{subfigure}
		\hfill
		\begin{subfigure}[b]{0.32\textwidth}
			\includegraphics[trim={0cm 0 0cm 0cm},clip,,width=\textwidth]{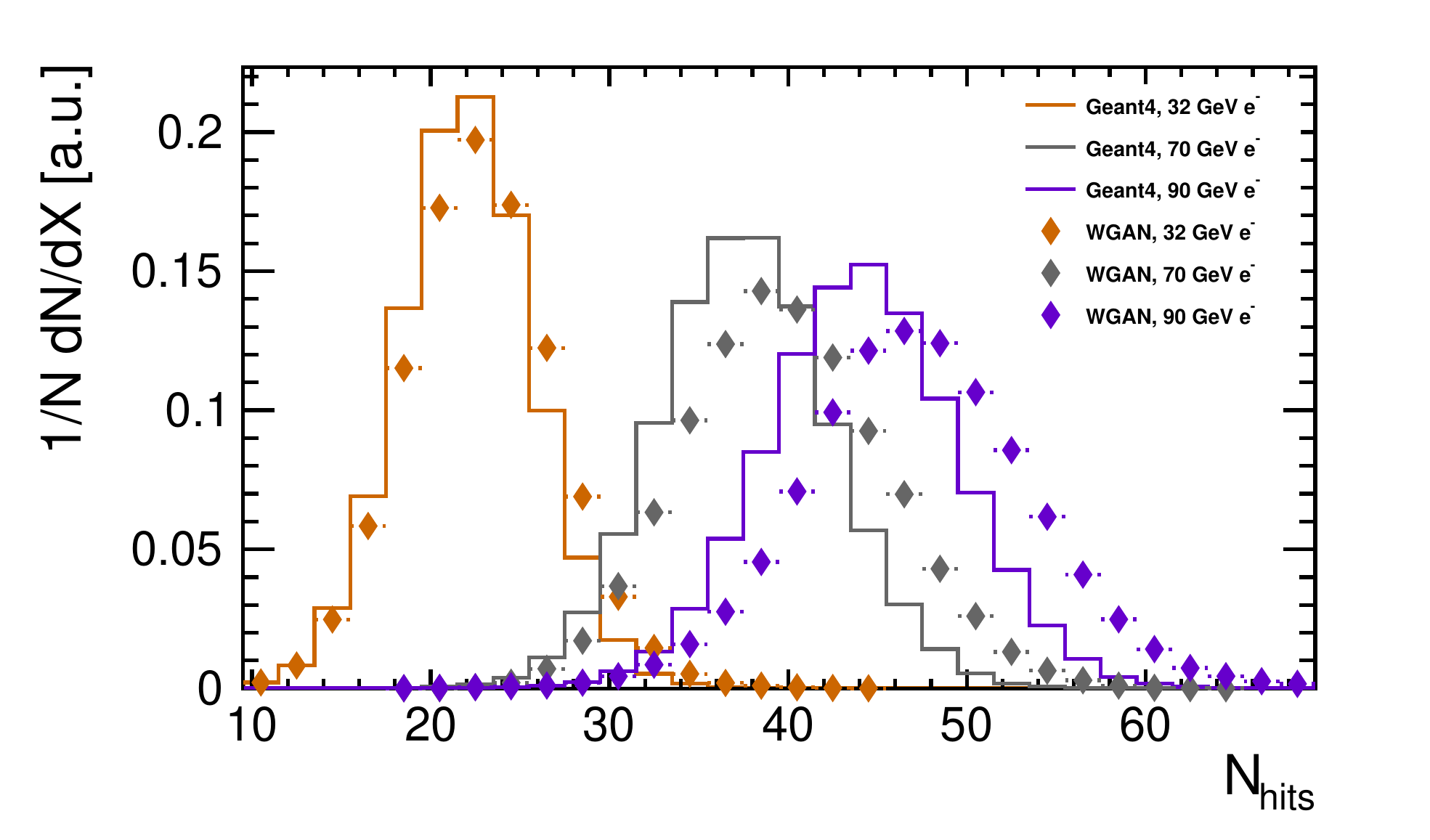}
			\subcaption{}

		\end{subfigure}	
		\hfill
		\begin{subfigure}[b]{0.32\textwidth}
			\includegraphics[trim={0cm 0 0cm 0cm},clip,,width=\textwidth]{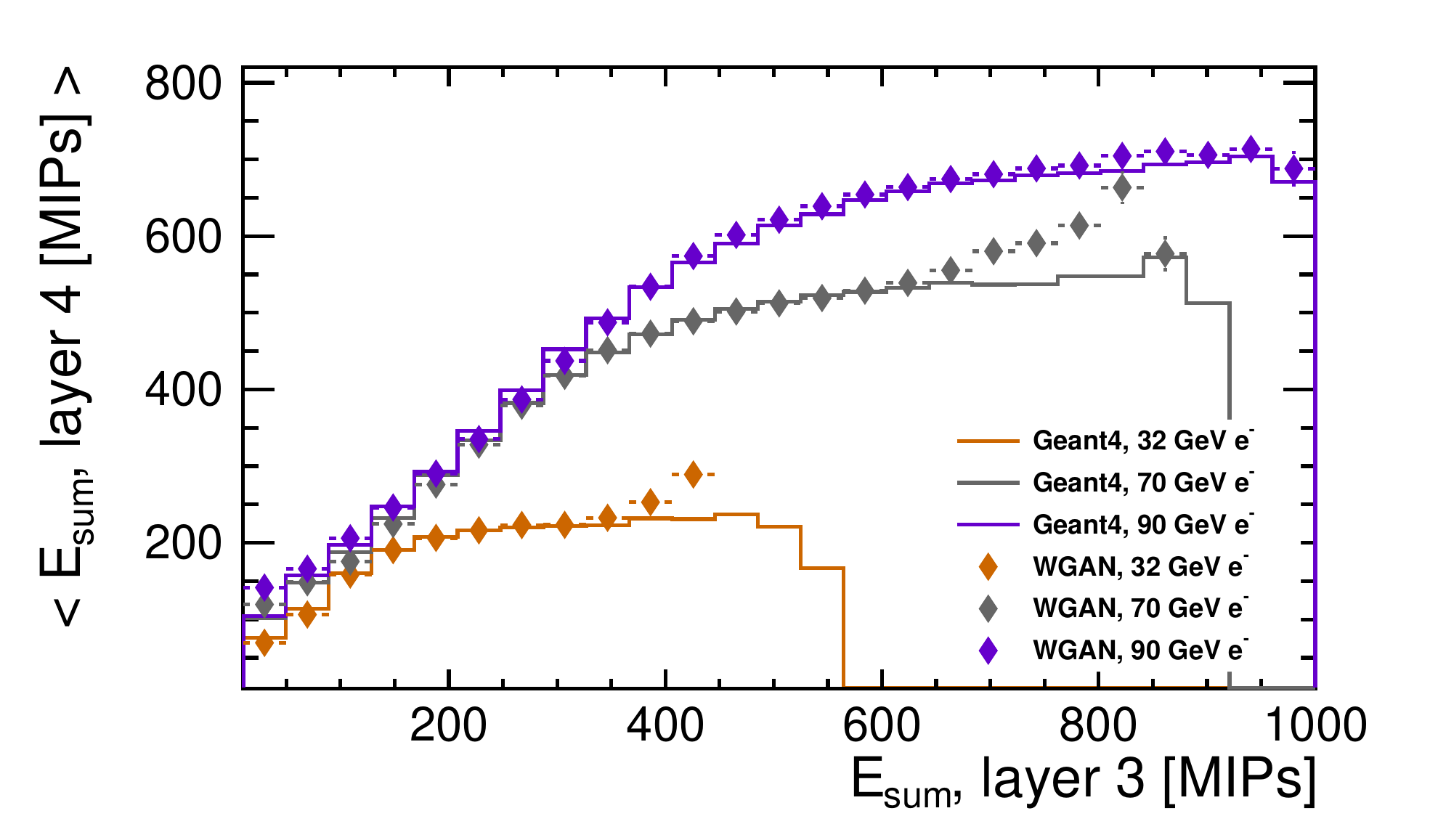}
			\subcaption{}
		\end{subfigure}
		\begin{subfigure}[b]{0.32\textwidth}
			\includegraphics[trim={0cm 0 0cm 0cm},clip,,width=\textwidth]{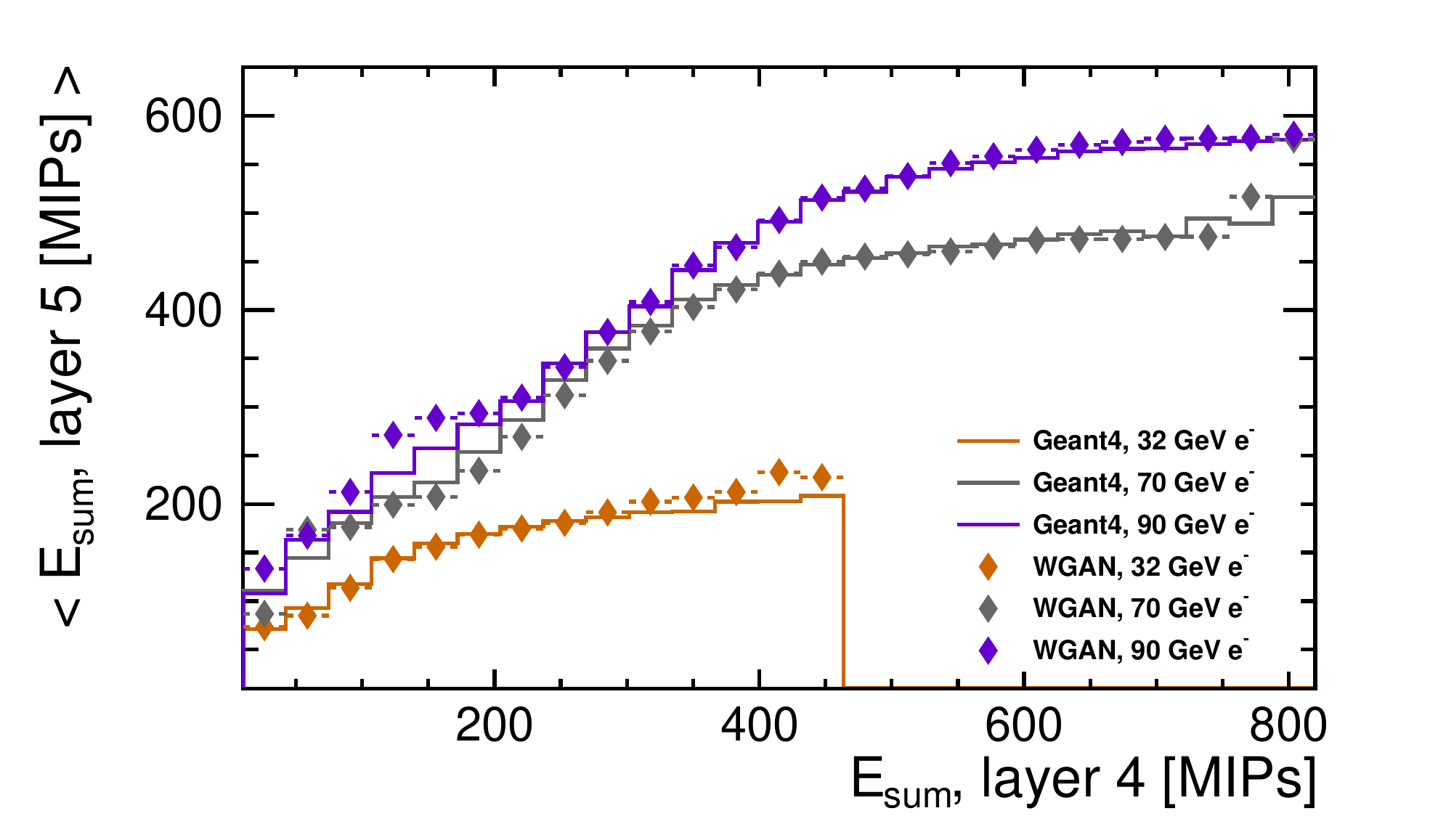}
			\subcaption{}
		\end{subfigure}
		\hfill		
		\begin{subfigure}[b]{0.32\textwidth}
			\includegraphics[trim={0cm 0 0cm 0cm},clip,,width=\textwidth]{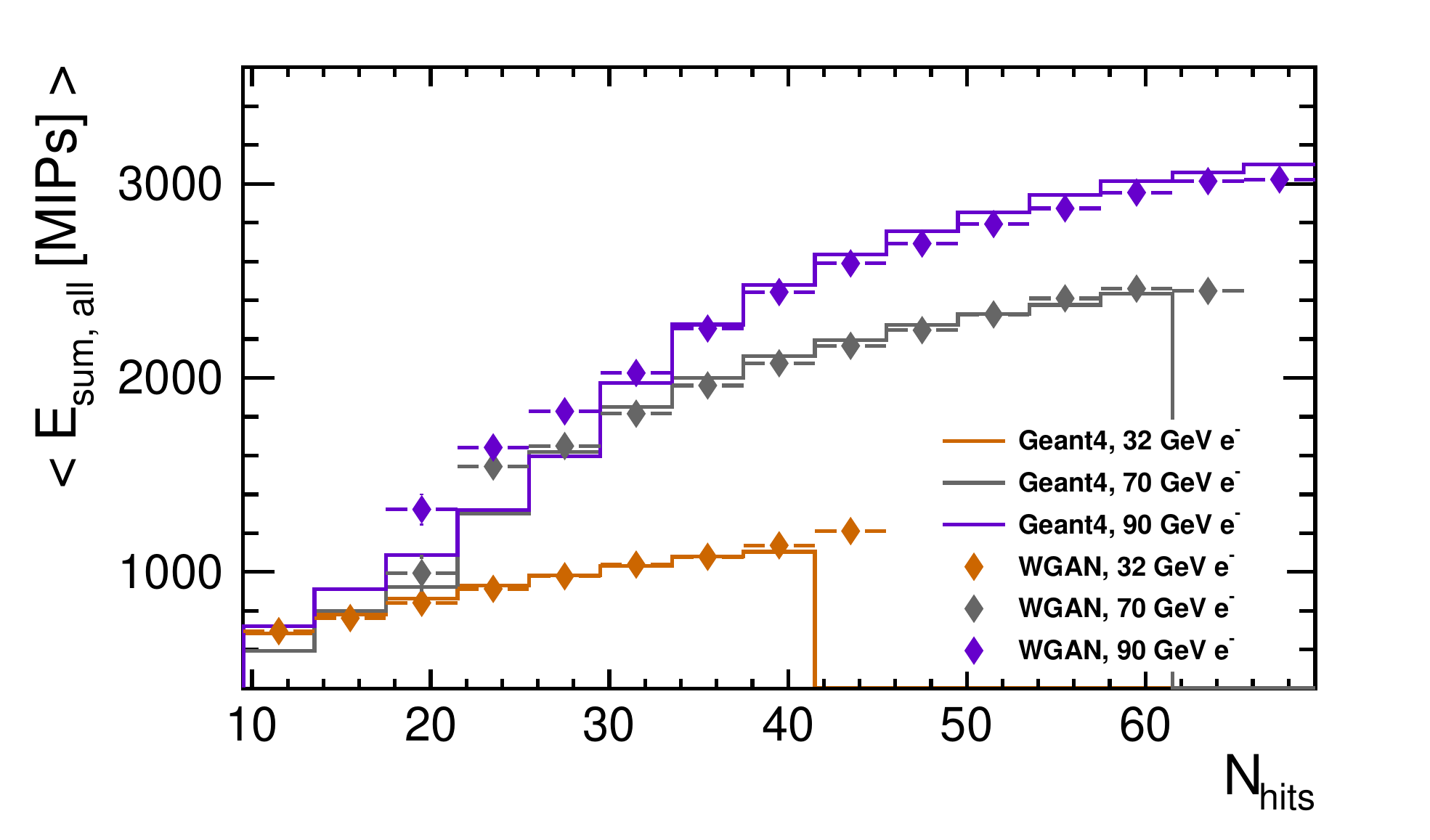}
			\subcaption{}
		\end{subfigure}
		\hfill
		\begin{subfigure}[b]{0.32\textwidth}
			\includegraphics[trim={0cm 0 0cm 0cm},clip,,width=\textwidth]{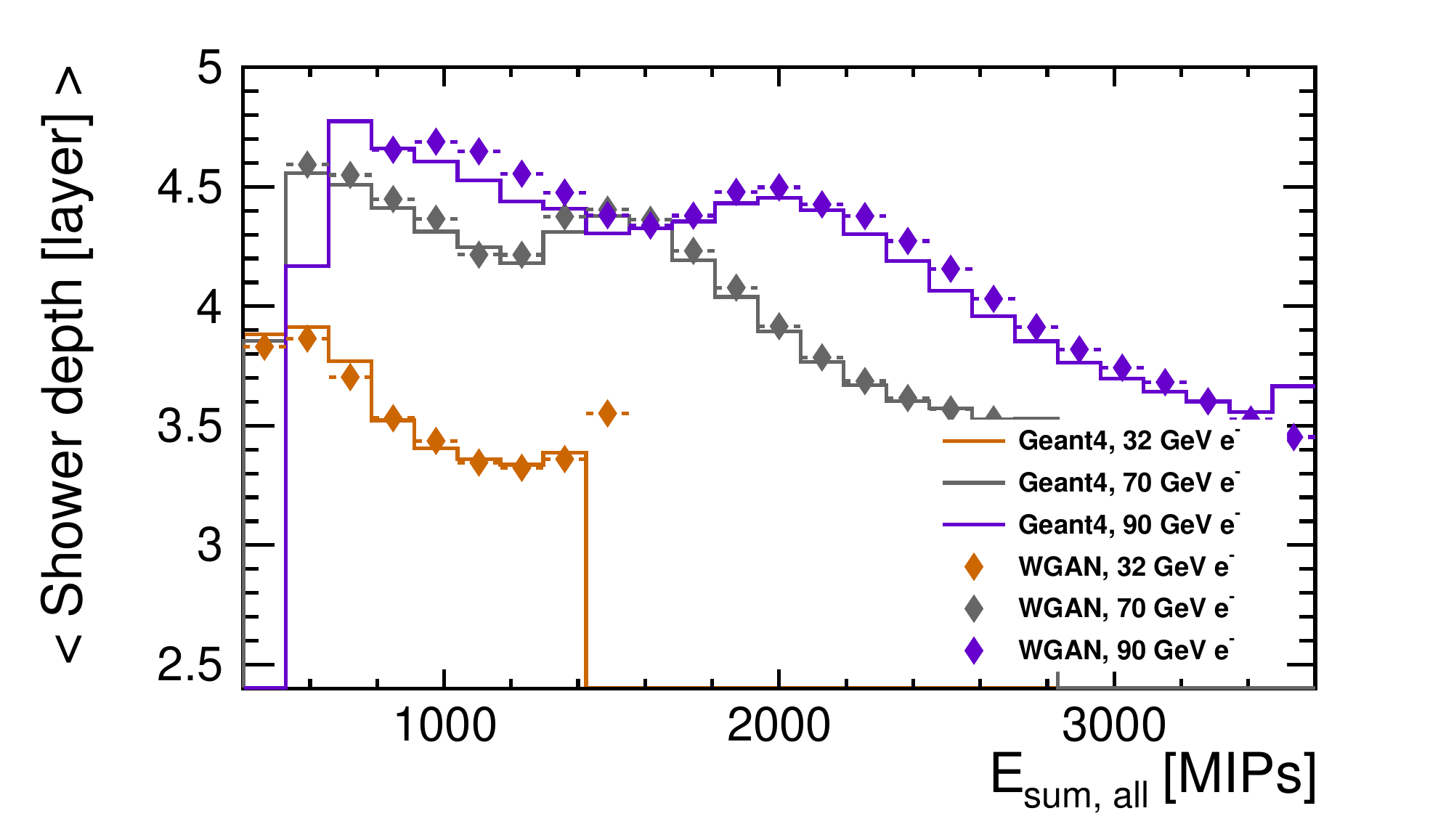}
			\subcaption{}
		\end{subfigure}
		\caption{Analogous to Figures \ref{figure:OneDComparison}, \ref{figure:NhitsComparison} and \ref{fig:correlations}. The noise threshold is set to 10 MIPs instead of 2 MIPs.}
		\label{fig:Appendixcorrelations}
\end{centering}
\end{figure}


\begin{thebibliography}{10}
\providecommand{\url}[1]{{#1}}
\providecommand{\urlprefix}{URL }
\expandafter\ifx\csname urlstyle\endcsname\relax
\providecommand{\doi}[1]{DOI~\discretionary{}{}{}#1}\else
\providecommand{\doi}{DOI~\discretionary{}{}{}\begingroup
	\urlstyle{rm}\Url}\fi		
		
\bibitem{Agostinelli:2002hh}
  S.~Agostinelli {\it et al.} [GEANT4 Collaboration],
  ``GEANT4: A Simulation toolkit,''
  Nucl.\ Instrum.\ Meth.\ A {\bf 506} (2003) 250.
  doi:10.1016/S0168-9002(03)01368-8

\bibitem{Allison:2006ve}
  J.~Allison {\it et al.},
  ``Geant4 developments and applications,''
  IEEE Trans.\ Nucl.\ Sci.\  {\bf 53} (2006) 270.
  doi:10.1109/TNS.2006.869826

\bibitem{Allison:2016}
  J.~Allison {\it et al.},
  ``Recent developments in Geant4,''
  Nucl.\ Instrum.\ Meth.\ A {\bf 835} (2016) 186.
  doi:10.1016/j.nima.2016.06.125

\bibitem{2014arXiv1406.2661G}
{I.~Goodfellow et al}: {Generative Adversarial Networks}.
\newblock \href{http://arxiv.org/abs/1406.2661}{arXiv:1406.2661 [stat.ML]}  (2014)

\bibitem{2016arXiv161207828S}
{A.~Shrivastava et al}: {Learning from Simulated and Unsupervised Images through
  Adversarial Training}.
\newblock \href{http://arxiv.org/abs/1612.07828}{arXiv:1612.07828 [cs.CV]}  (2016)

\bibitem{Hooberman:2017nips}
B.~Hooberman, et al.: {Calorimetry with Deep Learning: Particle
Classification, Energy Regression, and Simulation for High-Energy
Physics}.
\newblock Proc. Deep Learning for Physical Sciences Workshop at the 31st
Conf. Neural Information Processing Systems (NIPS), 2017, Long Beach,
United States

\bibitem{Paganini:2017hrr}
M.~Paganini, L.~de~Oliveira, B.~Nachman: {Accelerating Science with Generative
  Adversarial Networks: An Application to 3D Particle Showers in Multilayer
  Calorimeters}.
\newblock Phys. Rev. Lett. \textbf{120}(4), 042003 (2018)

\bibitem{Paganini:2017dwg}
M.~Paganini, L.~de~Oliveira, B.~Nachman: {CaloGAN}.
\newblock Phys. Rev. D \textbf{97}, 014021 (2018)

\bibitem{Erdmann:2018kuh}
 M.~Erdmann, L.~Geiger, J.~Glombitza, D.~Schmidt: {Generating and Refining Particle Detector Simulations Using the Wasserstein Distance in Adversarial Networks}, Comput Softw Big Sci (2018) 2: 4. 
 \newblock \href{https://doi.org/10.1007/s41781-018-0008-x}{https://doi.org/10.1007/s41781-018-0008-x} 


\bibitem{2017arXiv170107875A}
M.~Arjovsky, S.~Chintala, L.~Bottou: {Wasserstein GAN}.
\newblock \href{http://arxiv.org/abs/1701.07875}{arXiv:1701.07875 [stat.ML]}  (2017)

\bibitem{2017arXiv170400028G}
I.~Gulrajani, et al: {Improved Training of Wasserstein GANs}.
\newblock \href{http://arxiv.org/abs/1704.00028}{arXiv:1704.00028 [cs.LG]}  (2017)

\bibitem{2016arXiv161009585O}
A.~Odena, C.~ Olah, J.~Shlens: {Conditional Image Synthesis With Auxiliary
  Classifier GANs}.
\newblock \href{http://arxiv.org/abs/1610.09585}{arXiv:1610.09585 [stat.ML]}  (2016)

\bibitem{tensorflow}
M. Abadi et al: {TensorFlow: Large-scale machine learning on heterogeneous systems},
\href{https://arxiv.org/abs/1603.04467}{arxiv:1603.04467 [cs.DC]},
\href{https://www.tensorflow.org}{https://www.tensorflow.org}.

\bibitem{Apollinari:2016HLLHC}
G. Apollinari, et al., \href{https://cds.cern.ch/record/2116337/files/CERN-2015-005.pdf}{High-Luminosity Large Hadron Collider}, \emph{Technical Design Report CERN-2015-005.} (December 2015) 


\bibitem{CMS:2008CMSExperiment}
The CMS Collaboration, \href{http://iopscience.iop.org/article/10.1088/1748-0221/3/08/S08004/meta}{The CMS Experiment at the CERN LHC}, \emph{JINST \textbf{3} S08004}  (August 2008) 

\bibitem{Contardo:2018TechnicalReportHGCal}
D. Contardo, et al., \href{https://cds.cern.ch/record/2020886/files/LHCC-P-008.pdf}{The Phase-2 Upgrade of the
CMS Endcap Calorimeter}, \emph{Technical Design Report CERN-LHCC-2017-023.  CMS-TDR-019. ISBN 978-92-9083-459-5.}  (April 2018)

\bibitem{Martelli:2018HGCalOverview}
A. Martelli., \href{http://inspirehep.net/record/1620207?ln=en}
{The CMS HGCal detector for HL-LHC Upgrade}, \emph{arXiv:1708.08234v1 [physics.ins-det]}  (February 2018)


\bibitem{Jain:2017BeamtestSummary2016}	
S. Jain, \href{http://iopscience.iop.org/article/10.1088/1748-0221/12/03/C03011}{Construction and first beam-tests of silicon- tungsten prototype modules for the CMS High Granularity Calorimeter for HL-LHC}, \emph{JINST \textbf{12} C03011}  (March 2017) 

\bibitem{Quast:2018BeamtestSummary2017}
T. Quast, \href{http://iopscience.iop.org/article/10.1088/1748-0221/13/02/C02044/pdf}{Construction and beam-tests of silicon-tungsten prototype modules for the CMS High Granularity Calorimeter for HL-LHC}, \emph{JINST \textbf{13} C02044}  (Februrary 2018) 

\bibitem{Spanggaard:1998H2DWCs}	
J. Spanggaard, \href{http://cds.cern.ch/record/702443/files/sl-note-98-023.pdf}{Delay Wire Chambers - A Users Guide}, \emph{SL-Note-98-023.}  (1998)

\bibitem{Banerjee:2017CMSPhysicsLists}
T. Banerjee, \href{http://iopscience.iop.org/article/10.1088/1742-6596/898/4/042005/pdf}{Validation of Physics Models of GEANT4 using Data from CMS Experiment}, \emph{J. Phys.: Conf. Ser. \textbf{898}} 042005.  (2017) 

\bibitem{github}
\href{https://github.com/cms-sw/cmssw/tree/CMSSW\_10\_0\_X}{https://github.com/cms-sw/cmssw/tree/CMSSW\_10\_0\_X}  

\bibitem{ATLAS:2005NIMPR}
J. Colas et al [ATLAS Collaboration]: \href{http://inspirehep.net/record/683219/}{Position resolution and particle identification with the ATLAS EM calorimeter}.
\newblock Nuclear Instruments and Methods in Physics Research Section A \textbf{550}, Pages 96-115 (2005)

\bibitem{CMS:FastSim}
Soon Yung Jun, \href{http://stacks.iop.org/1742-6596/293/i=1/a=012023}{Gflash as a Parameterized Calorimeter Simulation for the CMS Experiment}, \emph{J. Phys.: Conf. Ser. \textbf{293}} 012023.  (2011) 

\end{thebibliography}
\end{document}